%% file: ms.tex
\documentclass[12pt,preprint,longabstract]{aastex}

\newcommand{\kms}{km s$^{-1}$}

\shorttitle{CO OTF Survey of the Virgo Cluster Spirals}
\shortauthors{Chung et al.}

\begin{document}

\title{$^{\textbf{12}}{\rm \textbf{CO(J=1-0)}}$ ON-THE-FLY MAPPING SURVEY \\
OF THE VIRGO CLUSTER SPIRALS. \\
I. DATA \& ATLAS}

\author{E. J. Chung\altaffilmark{1,3}, M.-H. Rhee\altaffilmark{2,3}, H. Kim\altaffilmark{3}, 
Min S. Yun\altaffilmark{4}, M. Heyer\altaffilmark{4}, and J. S.
Young\altaffilmark{4}}

\altaffiltext{1}{Dept. of Astronomy, Yonsei University, Seoul 120-749, Korea}
\email{rigelej@yonsei.ac.kr}
\altaffiltext{2}{Yonsei University Observatory, Yonsei University, Seoul
120-749, Korea}
\altaffiltext{3}{Korea Astronomy \& Space Science Institute, Daejeon 305-348,
Korea}
\altaffiltext{4}{Dept. of Astronomy, University of
Massachusetts, 710 North Pleasant Street, Amherst, MA 01003, USA}

\begin{abstract}
We have performed an On-The-Fly (OTF) mapping survey of ${\rm
^{12}{CO(J=1-0)}}$ emission in 28 Virgo cluster spiral galaxies
using the Five College Radio Astronomy Observatory (FCRAO) 14-m
telescope. This survey aims to characterize the CO distribution,
kinematics, and luminosity of a large sample of galaxies covering
the full extents of stellar disks, rather than sampling only the
inner disks or the major axis as was done by many previous single
dish and interferometric CO surveys. CO emission is detected in 20
galaxies among the 28 Virgo spirals observed. An atlas consisting of
global measures, radial measures, and maps,
is presented for each detected galaxy. A note
summarizing the CO data is also presented along with relevant
information from the literature. The CO properties derived from our
OTF observations are presented and compared with the results from
the FCRAO Extragalactic CO Survey by Young et al. (1995) which
utilized position-switching observations along the major axis and a
model fitting method. We find that our OTF derived CO properties agree 
well with the Young et al. results in many cases, but the Young et al. 
measurements are larger by a factor of 1.4 - 2.4 for seven (out of 18) cases.  
We will explore further the possible causes for the discrepancy in the 
analysis paper currently under preparation.
\end{abstract}

\keywords{atlases --- galaxies: clusters: individual (Virgo)
--- galaxies: ISM --- radio lines: galaxies}

\section{Introduction}

The high galaxy density and the proximity make the Virgo cluster a
particularly interesting laboratory for a galaxy evolution study.
Its dynamical evolution is still in progress, and evidence for
significant environmental effects is ubiquitous \citep[e.g.,
][]{chung07}. The nearness of the Virgo cluster makes it possible to
observe its member galaxies with excellent spatial resolution
($1\arcsec\ \sim 90$ pc). It is the first cluster for which
significant HI imaging was done \citep{vgo84}, and various new HI
surveys such as ALFALFA \citep{gio07} and VIVA \citep{chu07} have
recently been conducted.  The Virgo cluster has also been the
subject of many recent multi-wavelength surveys, such as in radio
continuum by NRAO VLA Sky Survey \citep[NVSS;][]{con98}, in
H$\alpha$ \citep{koo04,che06}, in UV by FAUST \citep{bro97} and
Galaxy Evolution Explorer (GALEX) \citep[e.g.,][]{dal07}, and in IR by Spitzer \citep{ken08}.
High quality optical images are also available from the Hubble Space 
Telescope (HST)/ACS
Virgo Cluster Survey \citep{cot04} and Sloan Digital Sky Survey
(SDSS).  

Here, we present the results from a new imaging survey of
$J=1\rightarrow0\ ^{12}$CO emission in a large sample of Virgo
galaxies in order to address the distribution and characteristics of
dense molecular gas in these galaxies. It is well established that
molecular clouds are the sites of ongoing star formation
(\citet{lar03} and references therein), and carbon monoxide (CO) is
the most commonly used tracer of molecular hydrogen (H$_2$), which
is the most abundant but invisible component of cold and dense
clouds \citep[e.g.,][]{sol91}. The global H$_2$ content and its
distribution in galaxies and a comparison with other gas and stellar
components as a function of morphological type, luminosity, and
environment are some of the key insights one can derive from CO
observations (see \citet{you91} and references therein).

Existing CO surveys of Virgo cluster galaxies suffer from limited
spatial coverage and small sample sizes.  The FCRAO extragalactic
survey \citep{young95} is the first CO survey covering a wide range
of distance, diameter, morphological type, and blue luminosity of
some 300 galaxies conducted using the FCRAO 14-m telescope,
including CO measurements (detections and upper limits) of 65 Virgo
galaxies. Because of the large observing time required, these
observations were conducted in the position-switching mode,
primarily along the optical major axis of the disks, and the global
CO line luminosity was derived assuming a model distribution. More
recently, high angular resolution CO images have been obtained using
interferometric measurements by \citet{sak99} and \citet{sof03}.
These measurements reveal a detailed molecular gas distribution at
100 pc scales, but they are limited only to the central region of
galaxies.  The BIMA SONG (Survey Of Nearby Galaxies; \citet{hel03})
has also carried out an imaging survey of CO emission in several
Virgo spiral galaxies, incorporating  the short spacing data from
the NRAO 12-m telescope. The total number of Virgo galaxies imaged
by the BIMA SONG is small (six), however.

These earlier CO observations have revealed important insights on
the molecular interstellar medium (ISM) in these galaxies. For example, CO emission is
often centrally concentrated, in contrast to the centrally deficient
HI distributions commonly found in these galaxies.  The molecular
gas distribution also shows little evidence for any influence of the
gas stripping mechanisms \citep[e.g.,][]{ken89}. No central CO peak
\citep{hel03} or nuclear molecular rings \citep[e.g.,][]{ion05} are
seen in other cases. The influence of cluster environment on the
molecular content is still poorly understood \citep[e.g.,][]{bos06}.

A distinguishing characteristic of our new CO survey is the
complete imaging of $^{\rm 12}$CO (J=1--0) emission of a large sample 
(28 galaxies) of Virgo spirals using the On-The-Fly (OTF) mapping mode of the Five College
Radio Astronomy Observatory (FCRAO) 14-m telescope. Our map size
of $10^{\prime} \times 10^{\prime}$ is larger than the optical 
diameter $\rm D_{25}$ and it covers the entire stellar disk of each galaxy.
The 45\arcsec\ angular resolution of the new CO images is well matched to the
existing VLA HI data and is well suited for comparison with other
high resolution multi-wavelength data. The specific questions we aim
to address are :

1) To what extent is the CO distribution governed by the disk
dynamics?

2) Is there a clear phase transition between HI and $\rm H_{2}$ as a
function of the interstellar radiation field?

3) Are there any systematic differences in the CO properties of spiral
galaxies in different environments?

We present in this paper the data and the CO atlas from our OTF
mapping survey. In section 2, we describe the sample selection.
Observations and data reduction process are described in Section 3.
The CO atlas and CO properties are presented in Section 4, and our
results are compared with those of \citet{young95}. Molecular gas
distribution in individual galaxies is discussed in Section 5, and
the summary and conclusion are given in Section 6. The
analysis and interpretation of the data addressing the above
questions will be presented in our subsequent papers.

\section{THE SAMPLE}

\begin{figure}
\epsscale{.80} \plotone{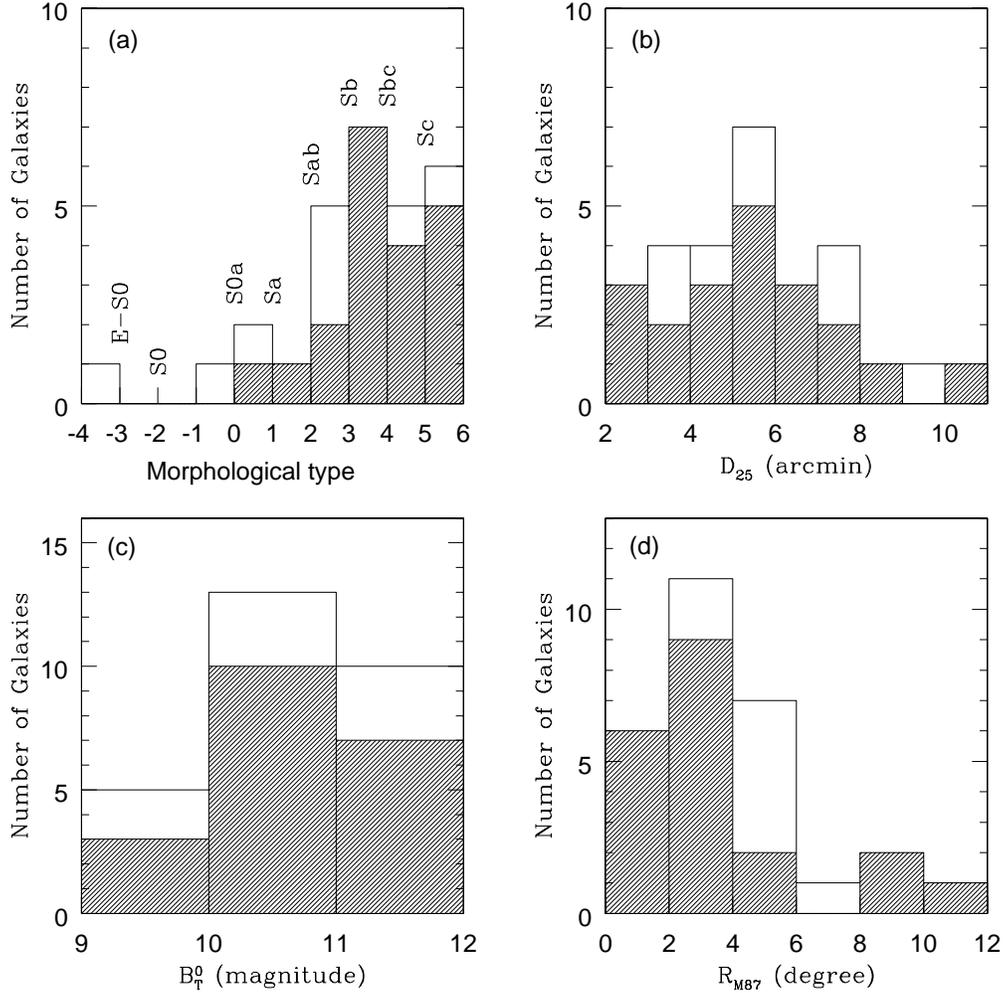}
\caption{The sample properties as a
function of (a) morphological type (LEDA), (b) major axis diameter
at 25th mag arcsec{$\rm ^{-2}$} in the $B$-band (LEDA), (c)
corrected total $B$-band magnitude (LEDA), and (d) angular distance
from M87. The white region represents the 28 Virgo galaxies
observed, and the shaded region corresponds to the 20 galaxies that
are detected in CO.\label{fig1}}
\end{figure}

We initially selected the 42 Virgo spiral galaxies in the magnitude limited 
sample studied by \citet{ken88}. The primary goal of our project is investigating 
the spatially resolved distribution of molecular gas and their relation to other 
tracers of activities among the Virgo spirals, and a large angular size 
is an important consideration. We also considered the total CO flux to increase the 
detection rate. Given the observing time limitations, a subset of 28 galaxies
was observed in the order of CO strength ($\rm S_{CO} \ge 200$ Jy \kms) and included 
in this study. 
The observed galaxies have an optical diameter of 3-10 arcminutes and span a wide 
range of morphology (Sa to Sc), surface brightness (9 $\lesssim \rm B_{T}^{0} 
\lesssim$ 12), and dust mass ($\rm 10^{6} \lesssim M_{dust} \lesssim 10^{7} M_{\odot}$). 
The basic properties of the observed galaxies are summarized in Table~1 and are
illustrated in Figure~\ref{fig1}.
\footnote{The E-SO galaxy NGC~4649 is included in our sample because
it is a companion of a late type spiral NGC~4647.}

% table 1 ==============================================
\clearpage
\input{tab1}
%============================================= end of table 1
%============================ Fig. 1 - the sample
%========================================== end of Fig. 1

\section{OBSERVATIONS \& DATA REDUCTION}

\subsection{Observations}

We carried out On-The-Fly (OTF) mapping observations of CO emission
in 28 Virgo galaxies over several observing sessions between January
2002 and February 2003 using the SEQUOIA (SEcond QUabbin Optical
Imaging Array) focal plane array receiver on the FCRAO 14-m
telescope. The SEQUOIA consists of 16 horns, each with
$45^{\prime\prime}$ beam size, configured in a $4\times4$ array. The
backend system used is the Quabbin Extragalactic Filterbanks (QEF),
which consists of sixteen independent spectrometers each with 64
channels at 5 MHz resolution, resulting in a total bandwidth of 320
MHz ($\Delta V\sim830$ \kms).  These spectra are calibrated using
the standard chopper-wheel method which corrects for atmospheric and
ambient temperature losses to yield the corrected antenna
temperature ${\rm {T_{A}}^{\ast}}$.

In the OTF mapping mode, the telescope moves fast and smoothly
across the target field taking the data continuously, and each map
pixel is sampled independently by all 16 independent detector
pixels.  Therefore, the use of the OTF observing mode offers a
significantly improved calibration, relative pointing accuracy and
registration, and a much higher dynamic range over the traditional
pixel-by-pixel mapping mode. Our OTF observations have fully covered
the entire stellar disk of each galaxy multiple times. Typically a
$10^{\prime} \times 10^{\prime}$ size box centered on each galaxy is
mapped with a scan speed of $45^{\prime \prime}$ per second, and the
data is stored in every 0.25 second. A reference spectrum is
obtained after every or every other row of scan at a location
30\arcmin\ away in the azimuth direction. A $6^{\prime}
\times 4^{\prime}$ region is mapped for NGC 4536 because of a
telescope problem during the observations.
NGC~4567 and NGC~4568 is an interacting pair observed simultaneously
using a single scanning box, and one data-cube contains both
galaxies.

The pointing and focus of the telescope are measured at 2-4 hour
intervals by observing the 86 GHz SiO maser in R-Leo. The measured
rms pointing error is $\sim 3^{\prime\prime}$ in both azimuth and
elevation, or $\sim 5^{\prime \prime}$ total.

\subsection{Data Reduction}
The data reduction is carried out in two steps, initially using the
revised-OTFTOOL \citep{chu06} and later the
GIPSY\footnote{\anchor{http:}{http://astro.rug.nl$/{\sim}$gipsy}}
(Groningen Image Processing SYstem; \citet{vdh92}) package. The
revised-OTFTOOL reads in the raw OTF data and produces a map  after
the initial editing and calibration. The resulting data cubes are
written out in FITS format and are imported to the GIPSY environment
for further data reduction and analysis.

The revised-OTFTOOL is a newly developed program based on the
OTFTOOL, which is the FCRAO facility pipeline software for the
SEQUOIA OTF data. The OTFTOOL was designed primarily for the
reduction of narrow Galactic emission line data taken with the
digital backend, and the default data filter is not well suited for
the baseline removal in the presence of weak, broad emission lines
seen in extragalactic CO data.  The revised-OTFTOOL includes several
new functions that are specifically designed to produce
noise-limited output images, with improved data filtering and
baseline fitting.  An improved filtering algorithm in the
revised-OTFTOOL identifies and removes bad spectra using the rms
level, antenna trajectory, elevation, and system temperature/gain
($\rm T_{sys}$), and data containing spikes are identified and
excluded. A second major improvement is the implementation of a new
self-referencing method. Rather than using the conventional ``OFF''
spectrum, our self-referencing method constructs the best OFF
spectrum from the OTF data itself by choosing the line-free regions
in the spatial and spectral domain (see Chung et al. 2005a \& 2005b
for more detail). This self-referencing method produces an OFF
spectrum temporally much closer to the ON spectra, significantly
reducing the influence of any residual gain changes and thus a
significantly improved baseline behavior. Finally, the
data are normally weighted and mapped onto a regular
$15^{\prime \prime}$ grid.

The GIPSY package is used to produce the final data cubes from the
individual scan maps.  Occasional bad filter bank channels are
removed and replaced with new data generated by interpolating two or
more adjacent channels.  Interpolation should produce a reasonable
result since the channel separation ($\Delta V \sim 13$ \kms) is
small compared with the intrinsic CO line width in these galaxies.
Data cubes including only the CO emission, used for further spectral
analysis, are created in two steps. First, a data mask for the CO
emission region is created through an iterative algorithm that
identifies the signal regions with criteria of 1.5 rms and
minimizes the noise in the line-free regions through a removal of a
low-order ($n\le2$) polynomial baseline in the frequency domain.
Then the final ''signal-only'' data cube is obtained from
the noise-minimized data cube by excluding the noise part and 
retaining the emission regions. For 6 galaxies which have large inclination
($\gtrsim 80$ degree) or weak emission(S/N ratio $\lesssim 2.5$),
Position-velocity diagram (PVD) is used to obtain physical quantities.\\

% table 2 =========================================================
\clearpage
\input{tab2}
% ============================================== end of table 2

\section{RESULTS}

Among the 28 galaxies observed, 20 galaxies including a galaxy pair
of NGC 4567 and NGC 4568 are detected in CO emission. The CO
detected galaxies are classified into two groups : (a) 14 galaxies with strong
emission features in the channel map with S/N ratio $\gtrsim
2.5$ (Group I); and (b) 6 galaxies with large inclination $\gtrsim 80$ degree
or weak CO emission with S/N ratio $\lesssim 2.5$ (Group II). We have produced
a CO atlas for these galaxies and derived their CO properties.
NGC~4567 and NGC~4568 are an interacting pair, poorly resolved by
our spatial and velocity resolutions. Therefore, their measured and
derived properties are reported as a single object. The results of NGC 4536
are derived from $6^{\prime} \times 4^{\prime}$ size data cube.\\
%================================== Figure 2~14
\clearpage
\begin{figure*}
\epsscale{0.6} \plotone{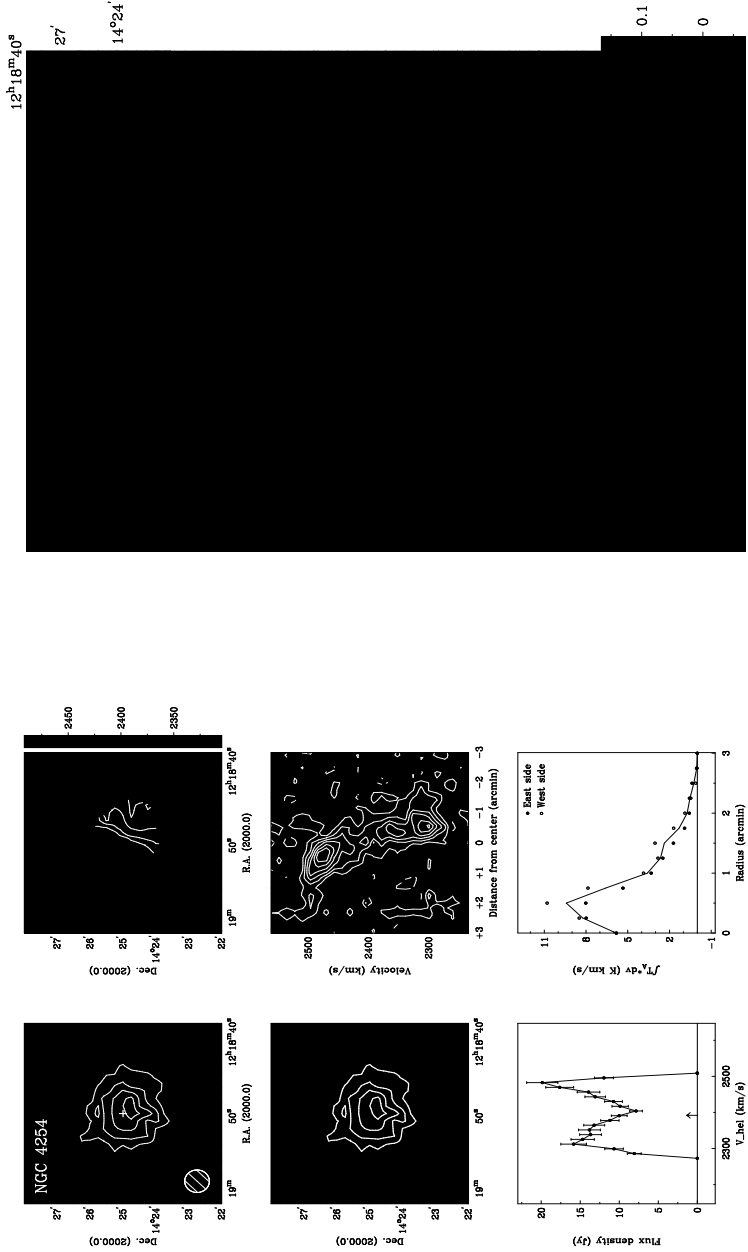} \caption{Group I-1. CO atlas and
channel maps of NGC~4254. CO intensity map (top left), velocity
field map (top right), CO contours overlapped on the optical
$B$-band image (middle left), position-velocity diagram
which is the central slice along the major axis of data
cube (middle right), global CO line profile (bottom left), and
radial CO distribution (bottom right). The line of radial
profile is the average of the east and west-side values for each
radial position. The channel maps are shown over the
velocity range of emission. The numbers on top right of the first
and second channels represent the velocities in km $s^{-1}$, and
their difference is the channel velocity separation.}
\end{figure*}
\clearpage
\begin{figure*}
\epsscale{0.6} \plotone{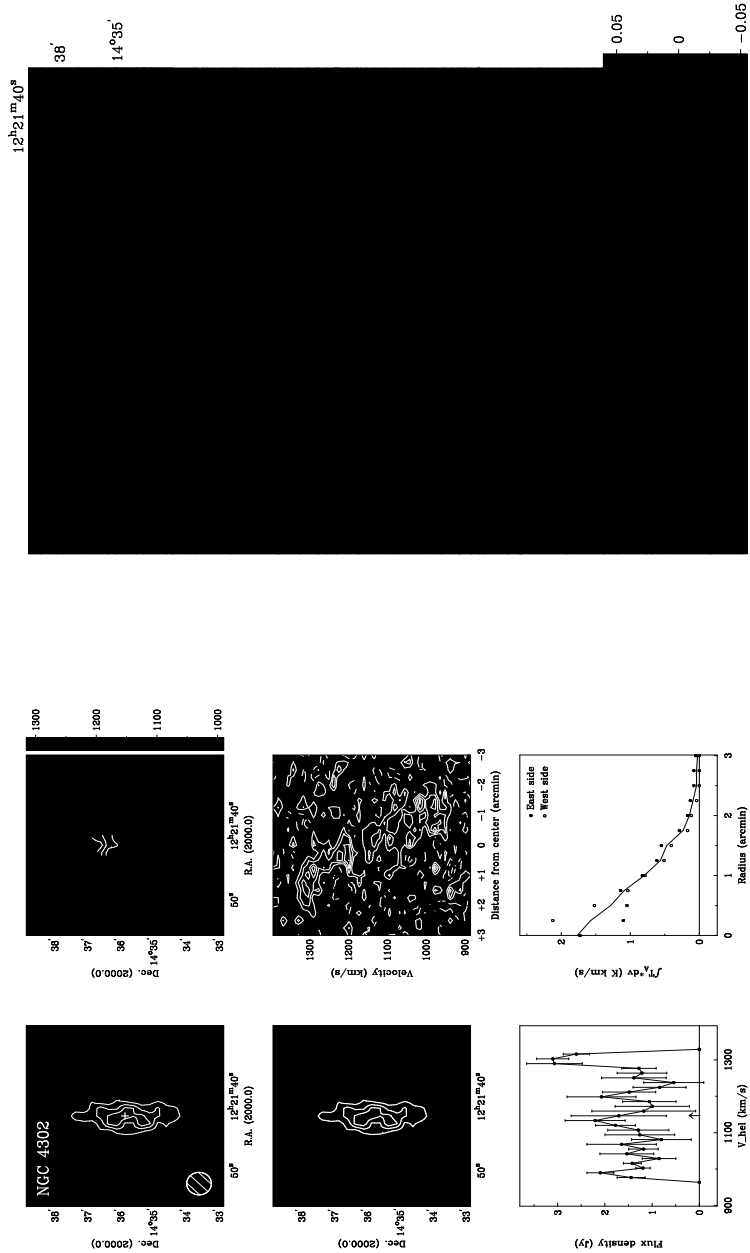} \caption{~Group I-2. Same as Figure
2 for NGC~4302. In the optical image, the companion galaxy NGC~4298
is visible in the east.}
\end{figure*} \clearpage
\begin{figure*}
\epsscale{0.6} \plotone{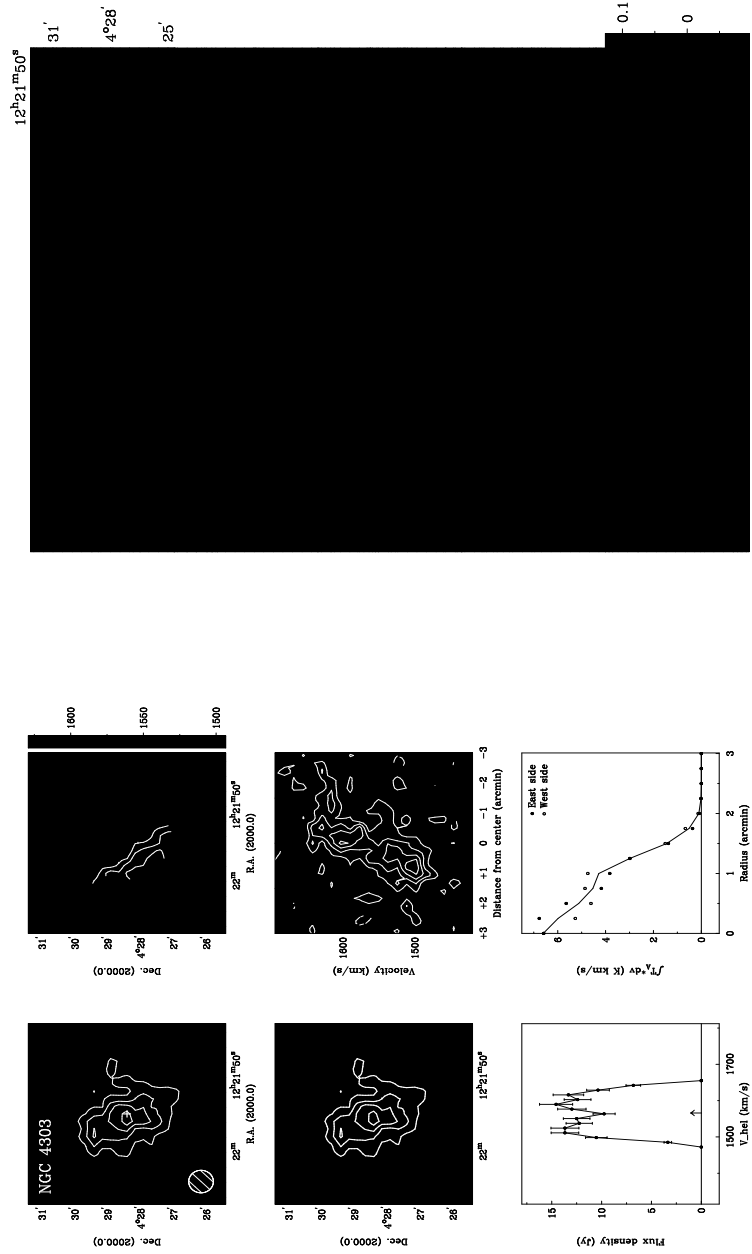} \caption{~Group I-3. Same as Figure
2 for NGC~4303.}
\end{figure*} \clearpage
\begin{figure*}
\epsscale{0.6} \plotone{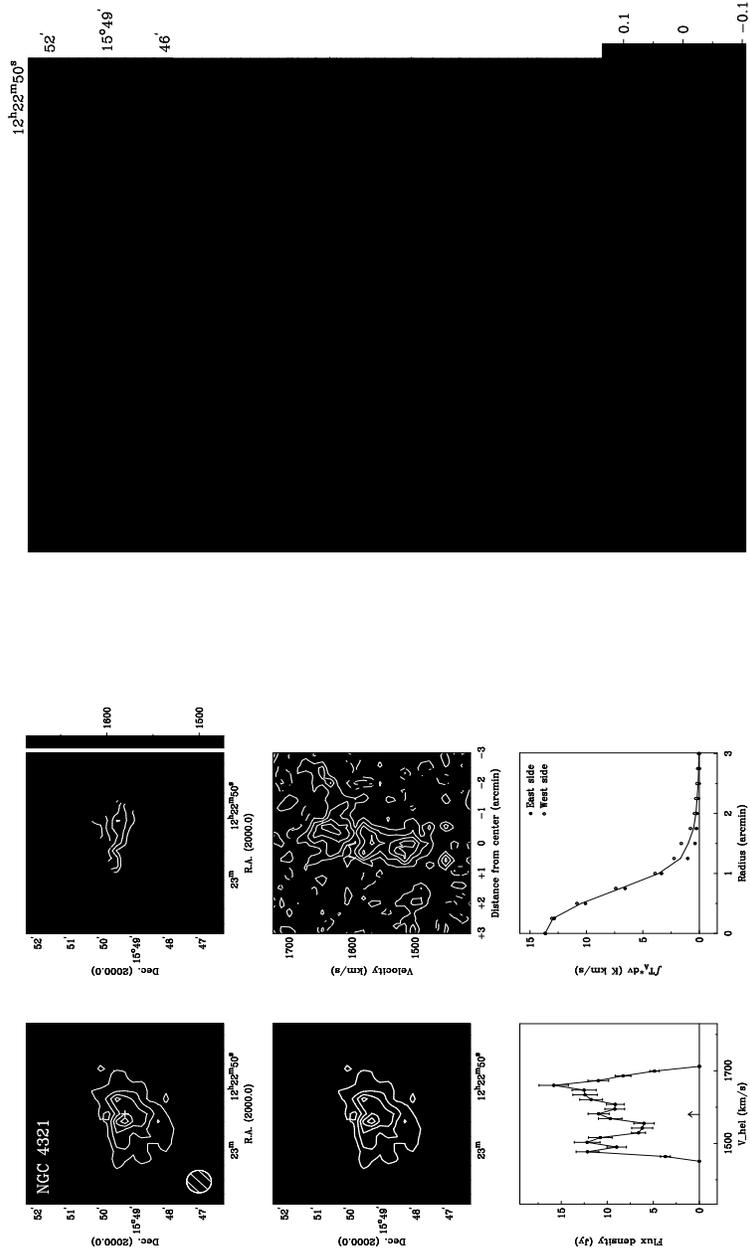} \caption{~Group I-4. Same as Figure
2 for NGC~4321.}
\end{figure*} \clearpage
\begin{figure*}
\epsscale{0.6} \plotone{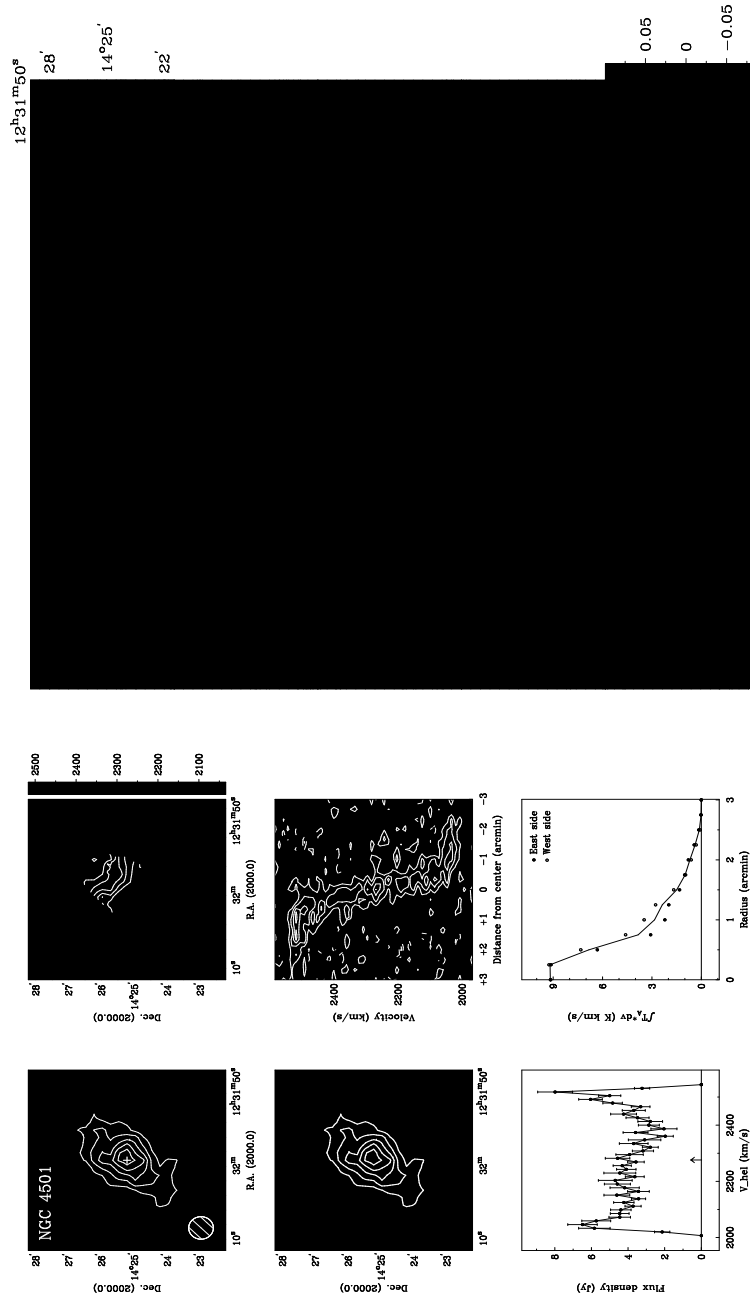} \caption{~Group I-5. Same as Figure
2 for NGC~4501.}
\end{figure*} \clearpage
\begin{figure*}
\epsscale{0.6} \plotone{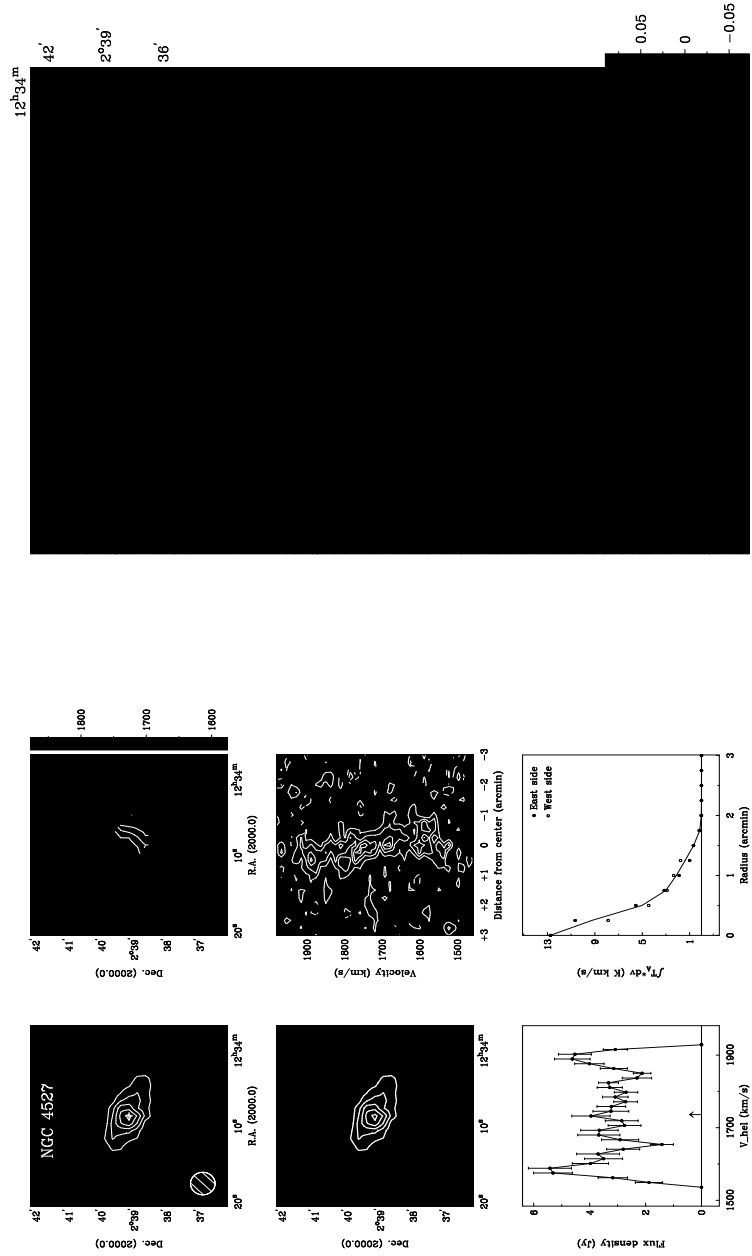} \caption{~Group I-6. Same as Figure
2 for NGC~4527.}
\end{figure*} \clearpage
\begin{figure*}
\epsscale{0.6} \plotone{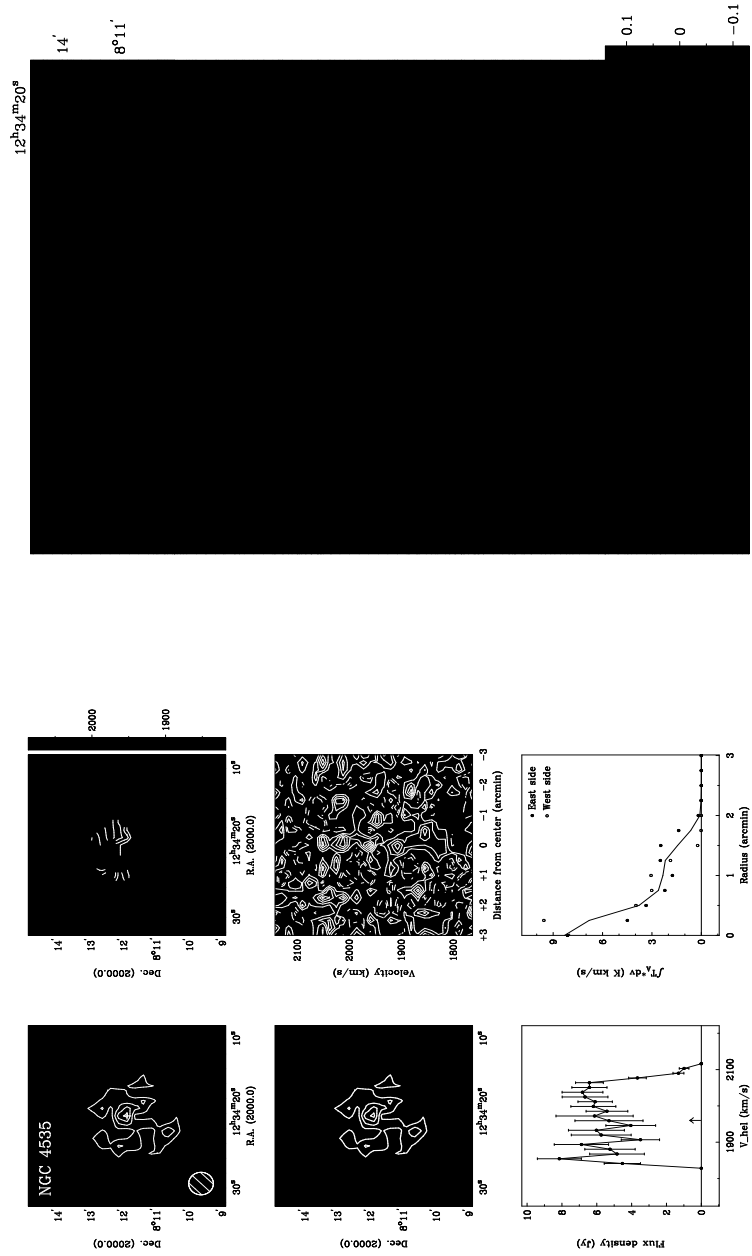} \caption{~Group I-7. Same as Figure
2 for NGC~4535.}
\end{figure*} \clearpage
\begin{figure*}
\epsscale{0.6} \plotone{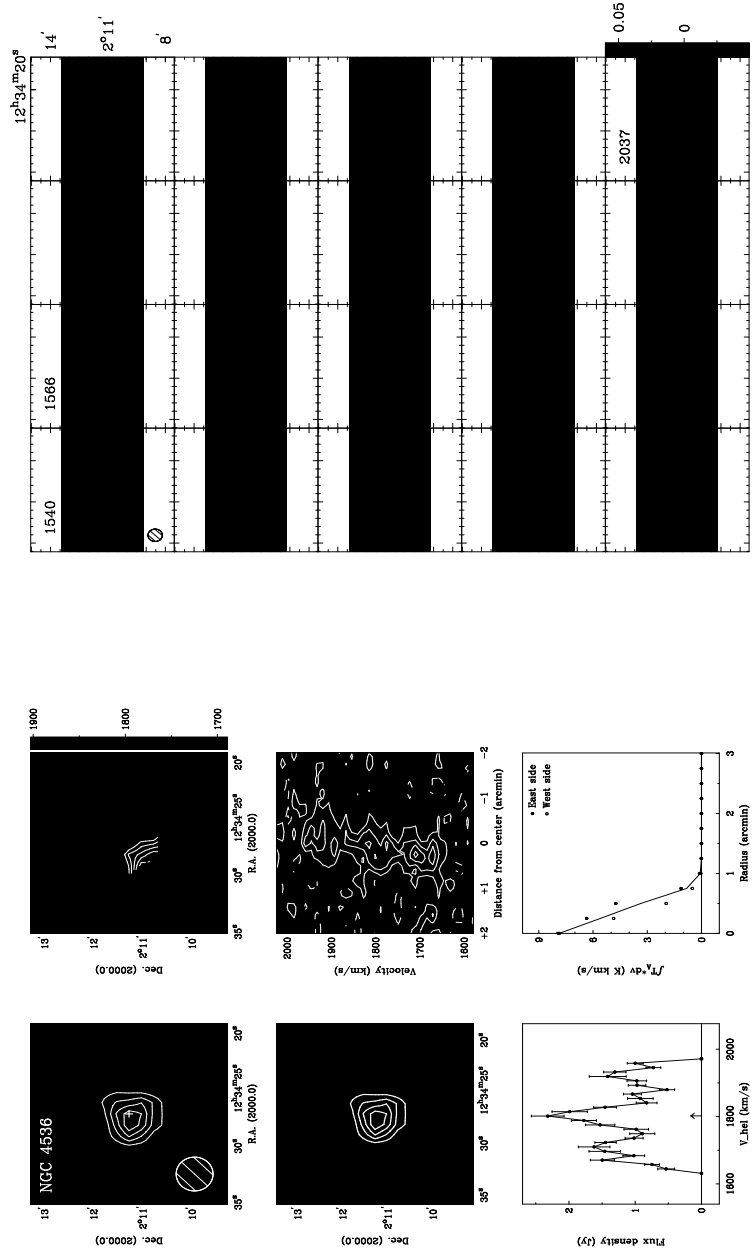} \caption{~Group I-8. Same as Figure
2 for NGC~4536.}
\end{figure*} \clearpage
\begin{figure*}
\epsscale{0.6} \plotone{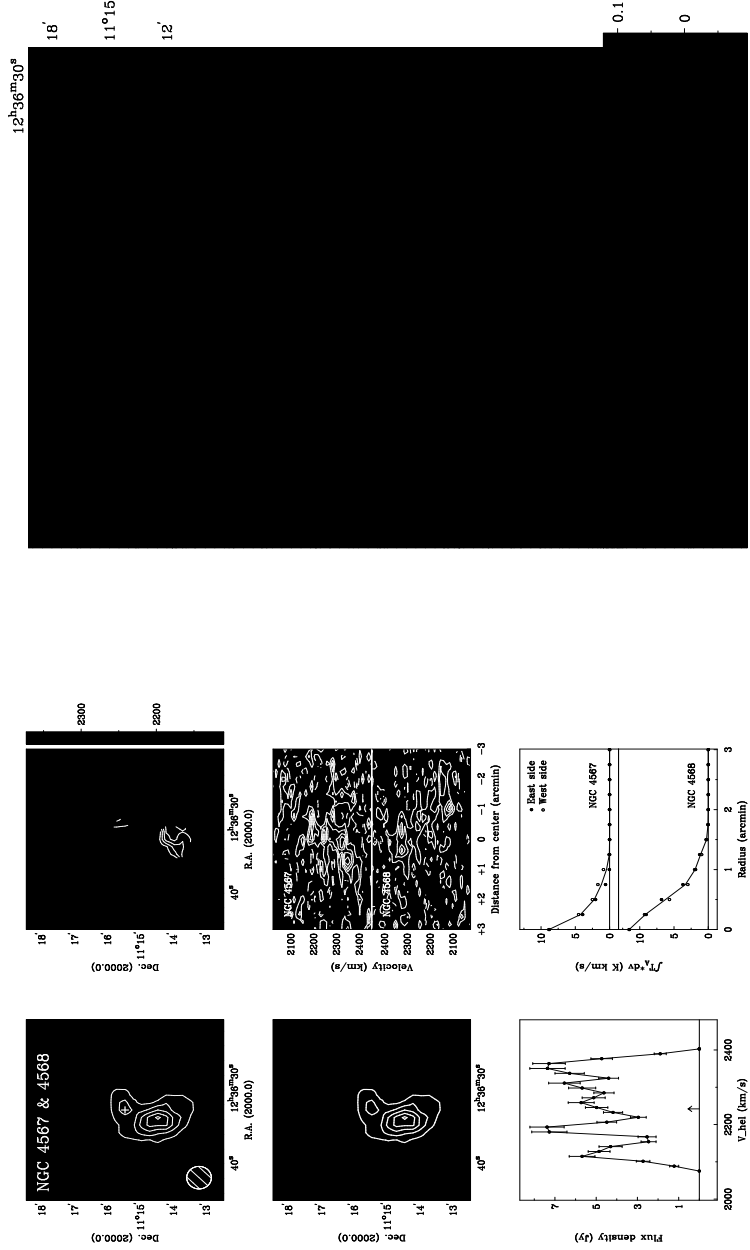} \caption{~Group I-9. Same as Figure
2 for NGC~4567 \& NGC~4568. NGC~4567 is in the center, and NGC~4568
is in the south. The PVDs and radial CO distributions are obtained
according to each galaxy's center and position angle.}
\end{figure*} \clearpage
\begin{figure*}
\epsscale{0.6} \plotone{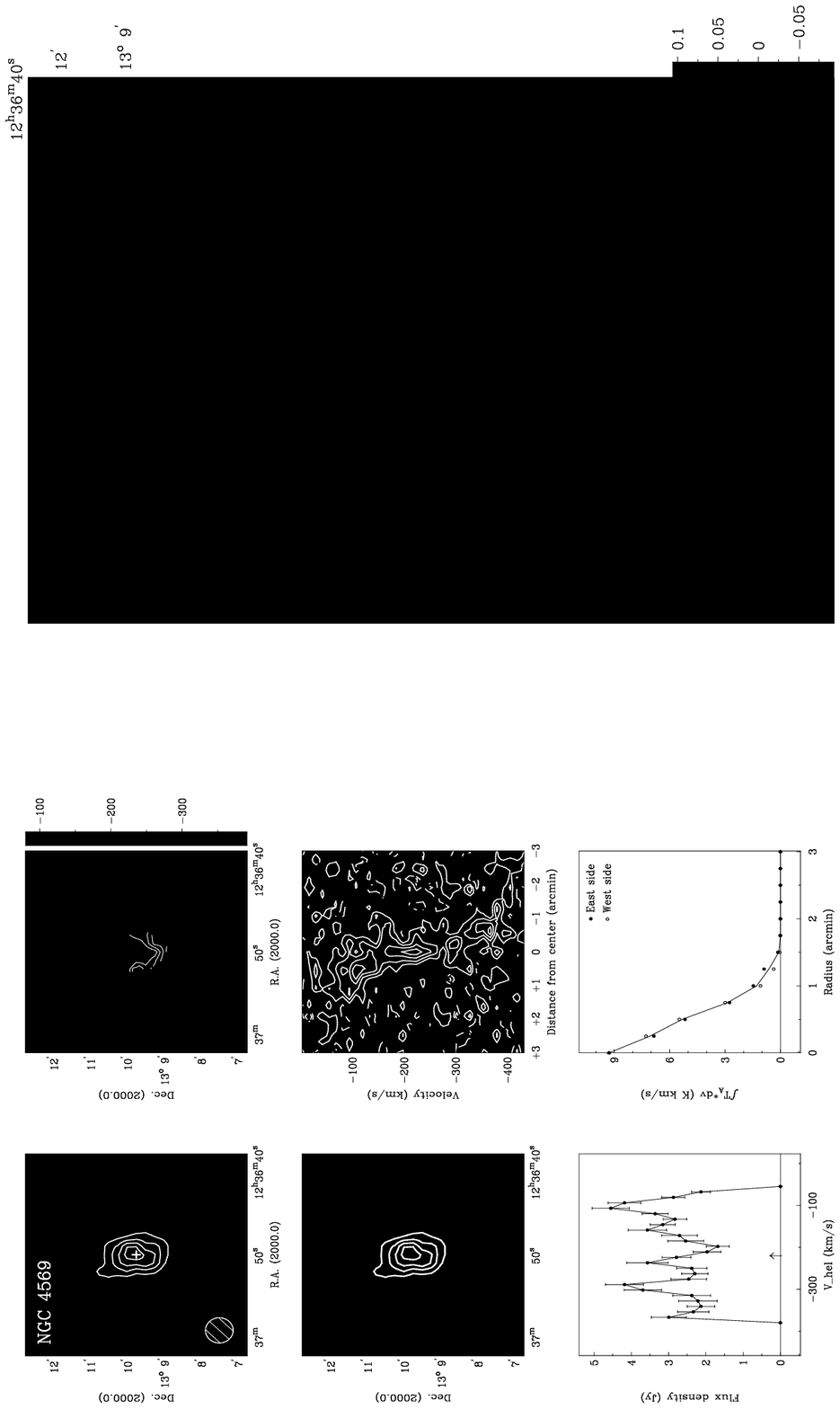} \caption{~Group I-10. Same as
Figure 2 for NGC~4569.}
\end{figure*} \clearpage
\begin{figure*}
\epsscale{0.6} \plotone{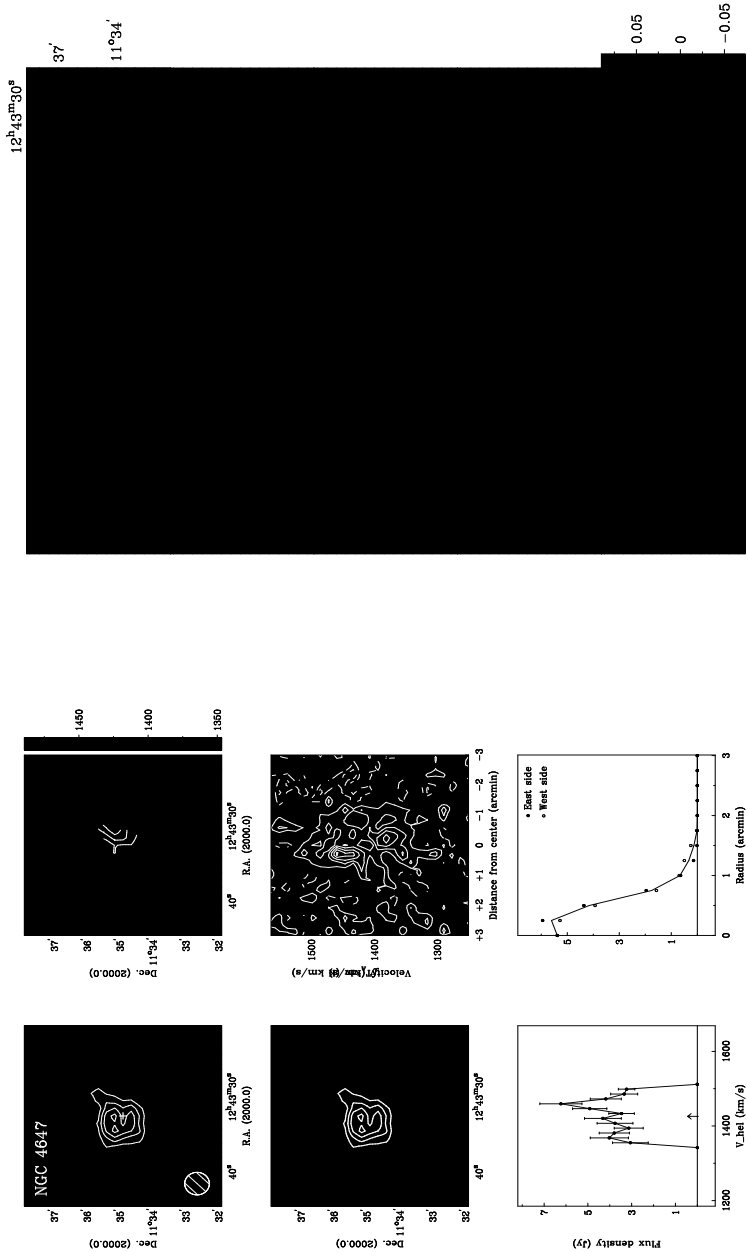} \caption{~Group I-11. Same as
Figure 2 for NGC~4647. In the optical image, the elliptical
companion galaxy NGC~4649 is visible in the east.}
\end{figure*} \clearpage
\begin{figure*}
\epsscale{0.6} \plotone{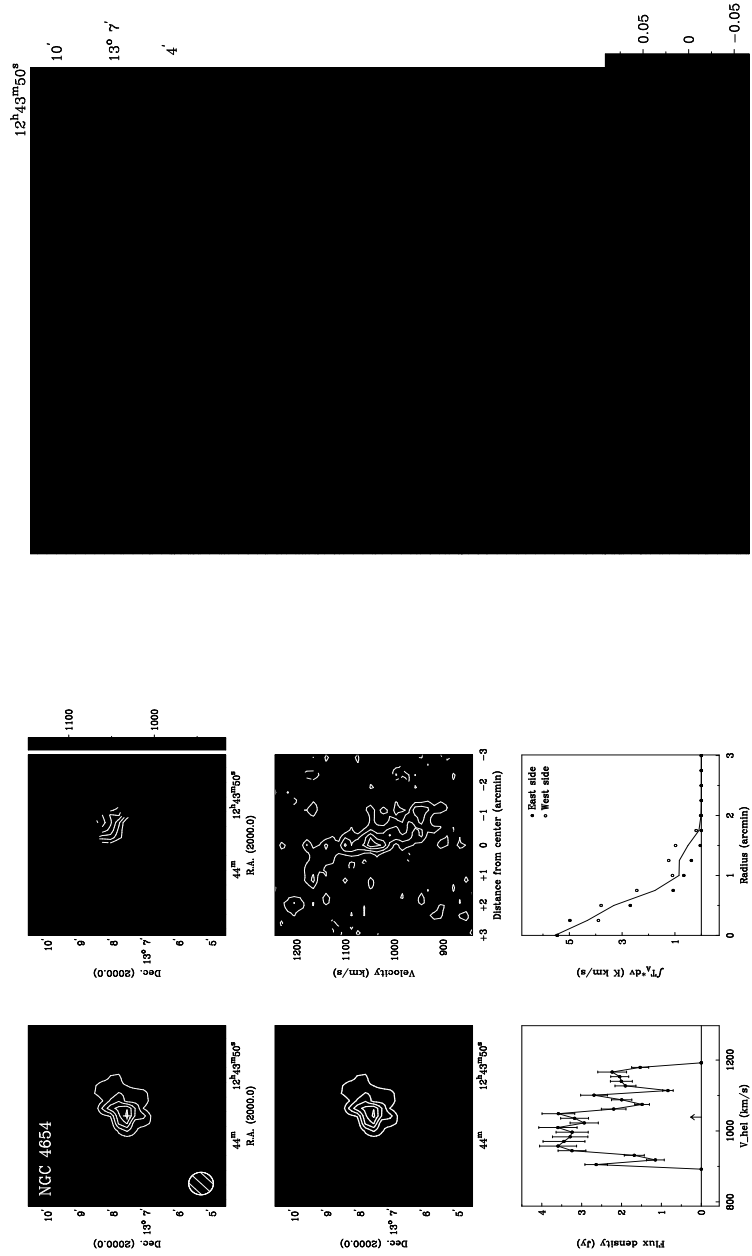} \caption{~Group I-12. Same as
Figure 2 for NGC~4654.}
\end{figure*} \clearpage
\begin{figure*}
\epsscale{0.6} \plotone{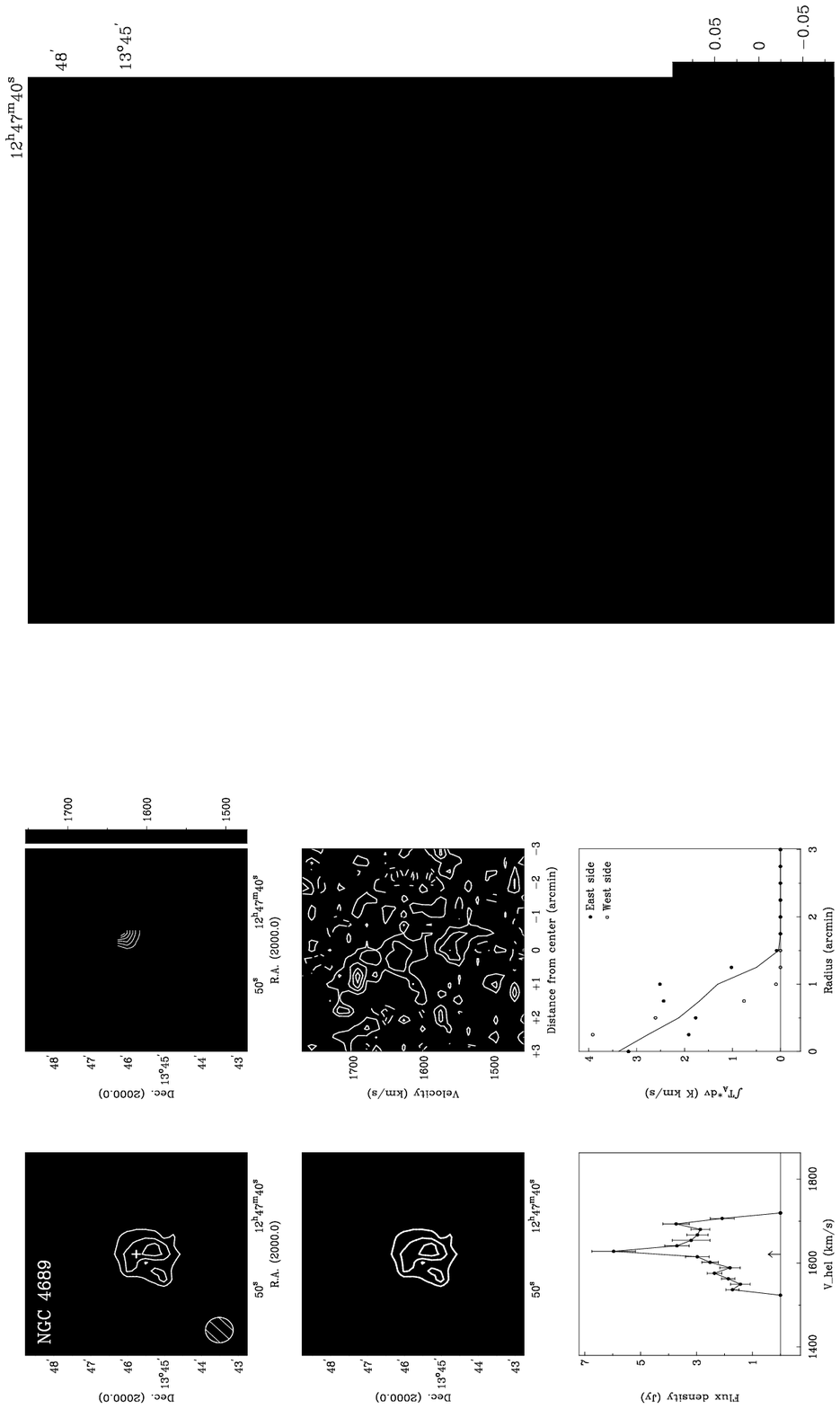} \caption{~Group I-13. Same as
Figure 2 for NGC~4689.}
\end{figure*}
%================================== end of figure 2~14
%================================== Figure 15~17
\clearpage
\begin{figure*}
\epsscale{0.45} \plotone{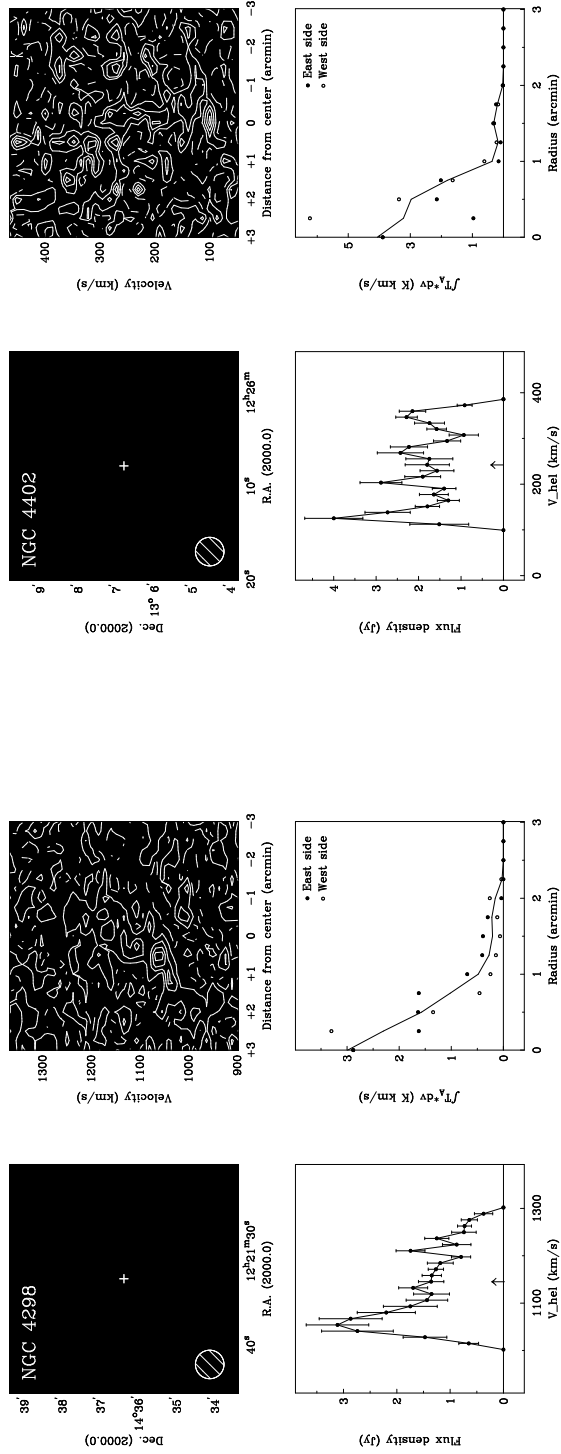} \caption{Group II-1. CO atlas of
NGC~4298(left) and NGC~4402(right). Optical $B$-band image (top
left), integrated position-velocity diagram (top right),
global CO line profile (bottom left), and radial CO distribution
(bottom right). In the optical image of NGC~4298, its companion
galaxy NGC~4302 is visible in the west.}
\end{figure*}
\clearpage
\begin{figure*}
\epsscale{0.45} \plotone{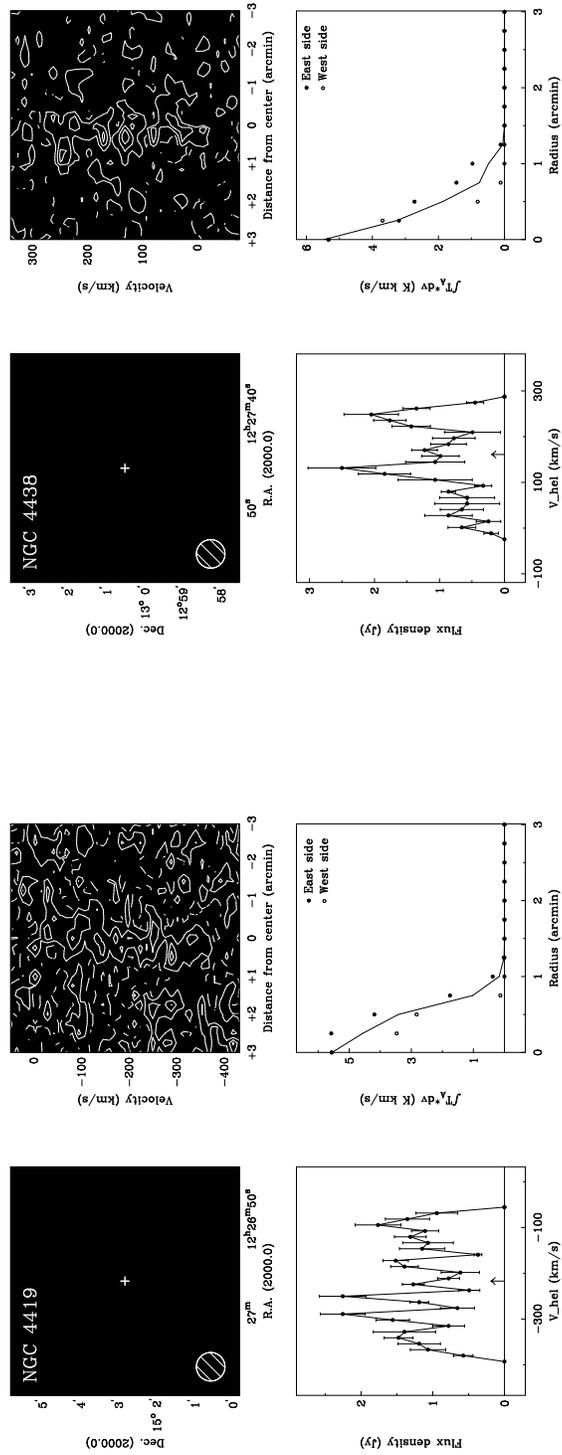} \caption{~Group II-2. Same as
Figure 15 for NGC~4419(left) and NGC~4438(right).}
\end{figure*}
\clearpage
\begin{figure*}
\epsscale{0.45} \plotone{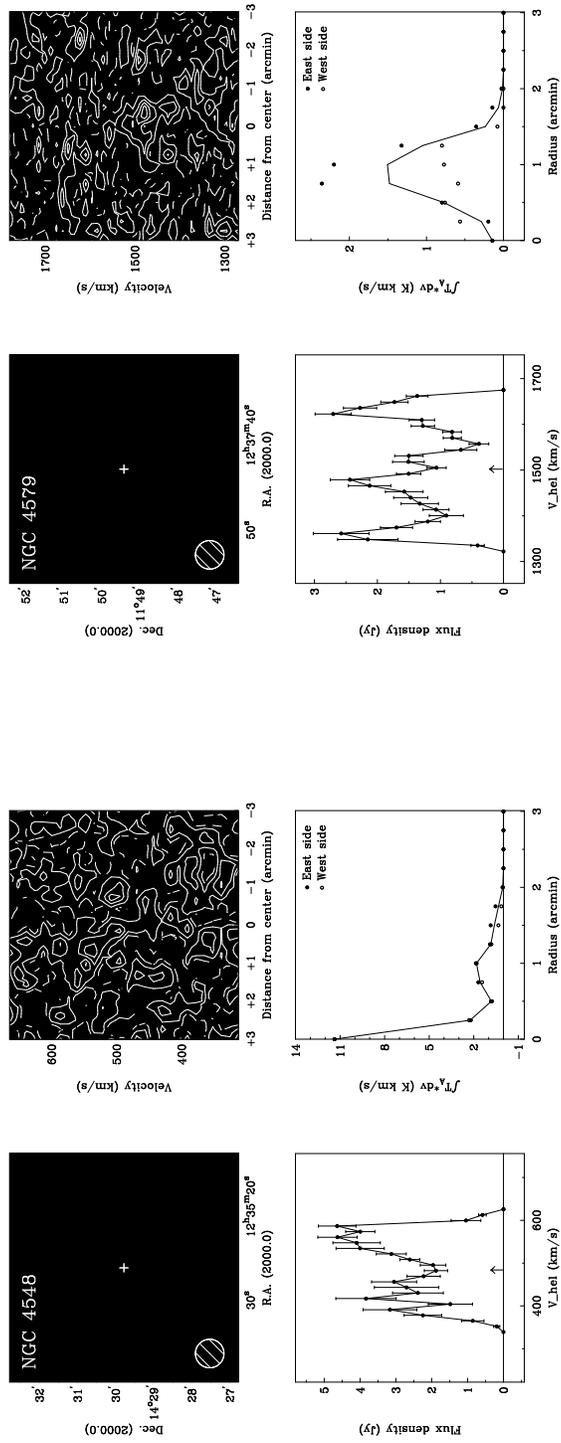} \caption{~Group II-3. Same as
Figure 15 for NGC~4548(left) and NGC~4579(right).}
\end{figure*}
%================================== end of figure 15~17

\subsection{CO Atlas: Map Descriptions} \label{bozomath}

Figures 2 -- 17 are the CO atlas and selected channel maps of Group
I and Group II galaxies obtained from our OTF mapping observations.
The atlas figure for each Group I galaxy consists of 6 maps: a CO
intensity map (top left), a velocity field map (top right), an
optical image with CO contours (middle left), a position-velocity
diagram derived along the major axis of the stellar disk (middle
right), a global CO line profile (bottom left), and a radial CO
profile (bottom right). Selected channel maps are also shown for
Group I galaxies. For Group II galaxies, CO emission is too
faint to yield an intensity map and velocity-field map. Therefore,
only an optical $B$-band image from NASA/IPAC Extragalactic Database
(NED\footnote{\anchor{http://nedwww.ipac.caltech.edu}{http://nedwww.ipac.caltech.edu}};
top left), a position-velocity diagram derived by integrating along
the minor axis (top right), a global CO line profile (bottom left),
and a radial CO profile (bottom right) are shown.

Each map shown (including the channel maps) has a dimension of
$6^{\prime}\times6^{\prime}$ ($4^{\prime}\times4^{\prime}$ for NGC
4536) since no emission was detected outside of this region for all
galaxies. The velocity range shown for the global CO profile and
channel map is about 500 km s$^{\rm -1}$, except for NGC~4501 which
has a CO line width of about 520 km s$^{\rm -1}$.
The contour levels shown are listed in Table 2.\\

\noindent \textbf{CO intensity map:} a CO intensity map is produced
by summing all emission features in the channel maps. The contour
levels shown are listed in K km s$^{\rm -1}$, and the CO
line intensity is in the T$_{\rm A}^*$ scale. \\

\noindent \textbf{Velocity field map:} a mean velocity field map is
derived through single Gaussian fit to the observed line profiles.\\

\noindent \textbf{Optical image:} An optical $B$-band image is
obtained from NED.\\

\noindent \textbf{Position-velocity diagram:} a PVD is extracted
from a row of pixels along the major axis, i.e., a central
slice of 15 arcsec width is shown in ATLAS,
for Group I galaxies. For Group II galaxies, whose inclination is too high
or whose CO emission is too weak in the individual channel maps, an integrated
PVD is produced by summing the data along the minor
axis. The integration is done over some minor axis length for
each galaxy (1 or 2 beam-width for highly inclined galaxies and up to 7 beam-width
for weak CO and face-on galaxies).\\

\noindent \textbf{Global CO line profile:} For Group I galaxies,
a global CO line profile is extracted from the data cube. For the
Group II galaxies, CO flux density is integrated manually along the
minor axis of the PVD, and the global line profile is computed
\citep[e.g.,][]{broe97,rhe05}. The uncertainty in the integrated line flux
in each channel is calculated as
\begin{equation}
\rm \sigma_{i}^{tot} = \sqrt{{\sigma_{i}^{cal}}^{2} +
{\sigma_{i}^{rms}}^{2} + {\sigma_{i}^{bsln}}^{2}}, \label{eq:equ1}
\end{equation}
\begin{equation}
\rm \sigma_{i}^{cal} = T_{i}^{\ast} \times 0.1, \label{eq:equ2}
\end{equation}
\begin{equation}
\rm \sigma_{i}^{rms} = T_{i}^{rms}, \label{eq:equ3}
\end{equation}
\begin{equation}
\rm \sigma_{i}^{bsln} = \sqrt{(T_{i}^{\ast}-T_{i0})^{2}+
(T_{i}^{\ast}-T_{i1})^{2}+(T_{i}^{\ast}-T_{i2})^{2}},
\label{eq:equ4}
\end{equation}
where ${\rm \sigma_{i}^{tot}}$ is the total uncertainty, ${\rm
\sigma_{i}^{cal}}$, ${\rm \sigma_{i}^{rms}}$, and ${\rm
\sigma_{i}^{bsln}}$ are the calibration uncertainty, thermal noise
in the channel maps, and the baseline subtraction noise. For the
calibration uncertainty, we assign a 10\% absolute uncertainty for
all spectra following \citet{ken88} (Eq. \ref{eq:equ2}). ${\rm
T_{i}^{\ast}}$ is CO line intensity, ${\rm T_{i}^{rms}}$ is map
noise level, and ${\rm T_{i0}}$, ${\rm T_{i1}}$, and ${\rm
T_{i2}}$ are CO intensities of the $i$-th channel derived by the
baseline subtraction of 0, 1, and 2 order, respectively
\citep[see][]{ken88}.

\noindent \textbf{Radial CO profile:} deriving a radial surface
density profile using a concentric ellipse fitting method does not
work well for edge-on or highly inclined galaxies.
\citet{war88} applied Lucy iteration method \citep{luc74} to obtain
HI radial surface density profile of edge on galaxies, and we
derived radial CO profiles following \citet{war88}.\\

\noindent \textbf{Channel map:} The velocity range of each channel
map is about 500 km s$^{\rm -1}$ (740 km s$^{\rm -1}$ for NGC~4501),
which is the same as that of the global CO line profile. \\

\subsection{CO Properties}

CO properties for the 20 CO detected Virgo spiral galaxies are
summarized in Table~3.  The column entries are: (1) NGC number; (2)
rms noise measured from emission-free regions; (3) line width
measured at 20\% level; (4) line width at 50\% level; (5) systemic
velocity derived from the CO velocity profile; (6) total CO line
flux; (7) molecular hydrogen mass; (8) effective CO diameter;
and (9) isophotal CO diameter.\\

\noindent \textbf{Total CO flux \& H$_2$ mass}

\noindent To convert the CO intensity in ${\rm T_A^*}$ into the flux
density unit, a calibration factor of 42 Jy K$^{\rm -1}$
obtained using the FCRAO 14-m
telescope \citep{ken88} is applied. We derive ${\rm H_{2}}$ masses from
total CO flux measured assuming a linear conversion relation
\citep{ken89},

\begin{equation} \label{eq:mh2}
\rm M_{H_2} = 3.9 \times 10^{-17} \chi d^{2} S_{CO} (M_{\odot})
\end{equation}

\noindent where $\chi$ is the conversion factor and $d$ is distance
to the source in megaparsec (Mpc). We adopt $\chi = 3 \times
10^{20}$ ${\rm cm}^{-2}$ ( K[T$_{\rm R}$] km s$^{\rm -1}$ )$^{\rm
-1}$ \citep{you91}. The distance to the Virgo cluster is somewhat
uncertain because of the large depth effect \citep{yas97} and is
generally thought to be between 15 and 20 Mpc
\citep[e.g.,][]{young95,sak99,sof03,san06,mei07}. We adopt a
distance of 20 Mpc for an easier comparison with the FCRAO
Extragalactic CO Survey (see \S~\ref{sec:comparison}).\\

\noindent \textbf{CO line width}

\noindent CO line width is measured at 20\% and 50\% level of the
line peak, on each side of the line, following the definition of
\citet{rhee96}. The line profile is divided into two equal velocity
bins, and the peak fluxes ${\rm T_{low}^{peak}}$ and ${\rm
T_{high}^{peak}}$ are determined separately on each half of the line
profile.  The 20\% and 50\% velocities ($V^{\rm 20\%}$ and $V^{\rm
50\%}$) represent the velocities at which the line profile reaches
the 20\% and 50\% of the peak value on the respective high and low
velocity side, approaching from the line edge to the center of the
line profile. A linear interpolation procedure is used in this
calculation.

The final value of line width of each cutoff--level is determined as

\begin{equation}
\rm W_{20}^{obs} = V_{\rm high}^{\rm 20\%} - V_{\rm low}^{\rm 20\%}
\end{equation}
\begin{equation}
\rm W_{50}^{obs} = V_{\rm high}^{\rm 50\%} - V_{\rm low}^{\rm 50\%}.
\end{equation}

The uncertainty in the line width is estimated with 1 $\sigma$
uncertainty in each case where $\sigma$ is a mean rms noise of
line-free channels \citep{rhe96}.

The line width is corrected for instrumental broadening following the
method described by \citet{ver97}.  The correction for instrumental
broadening (in \kms) is computed as

\begin{equation}
\rm W_{20}=W_{20}^{\rm obs}-35.8 \left[\sqrt{1+\left({ \triangle V
\over 23.5}\right)^2}-1 \right] \label{eq:ibcl2}
\end{equation}
\begin{equation}
\rm W_{50}=W_{50}^{\rm obs}-23.5 \left[\sqrt{1+\left({ \triangle V
\over 23.5}\right)^2}-1 \right] \label{eq:ibcl5}
\end{equation}

\noindent where $\triangle V$ is the velocity resolution in
{km s$^{\rm -1}$}.
\\

\noindent \textbf{CO systemic velocity}

\noindent Again, following the procedure described by \citet{ver97},
we compute the systemic velocity of each galaxy as
\begin{equation}
\rm V_{\rm sys} = \left(V_{\rm low}^{20\%}+V_{\rm
high}^{20\%}+V_{\rm low}^{50\%} +V_{\rm high}^{50\%}\right) / 4
\label{eq:vsys}.
\end{equation}

\noindent \textbf{CO diameters}

\noindent To describe CO extents, we derived effective diameters and
isophotal diameters. Effective CO diameter, $\rm D_{CO}^{eff}$, is
defined as a diameter which encloses 70\% of the total emission
\citep[see][]{young95}. The CO isophotal diameter, $\rm
D_{CO}^{iso}$, is defined as a diameter where the mean face--on
surface density of $\rm H_{2}$ falls %into 
to 1 $\rm M_{\odot} pc^{-2}$
which corresponds to \rm{$\int T_{R}$} dv = 0.21 K \kms, or
\rm{$6.3 \times$ $10^{19}$} $\rm H_{2}$ $\rm cm^{-2}$ for a CO--$\rm H_{\rm 2}$
proportionality factor of \rm{$\chi = 3 \times$ $10^{20}$} $\rm cm^{\rm -2}$
(K[T$_{\rm R}$] km s$^{\rm -1}$ )$^{\rm -1}$
\citep{you91}.

% table 3 ========================================================
\clearpage
\input{tab3}
%============================================================== end of table 3

\subsection{Comparison With Other Observations \label{sec:comparison}}

Previous single dish and interferometric CO measurements exist for
many of our target galaxies.  A comparison of our results with those
of the FCRAO Extragalactic CO Survey \citep{young95} and the BIMA
SONG \citep{hel03} provides an important test of potentially
important systematics associated with these two widely utilized
extragalactic CO surveys.\\

\subsubsection{Comparison With the FCRAO Extragalactic CO Survey}

\begin{figure*}
\epsscale{0.8} \plotone{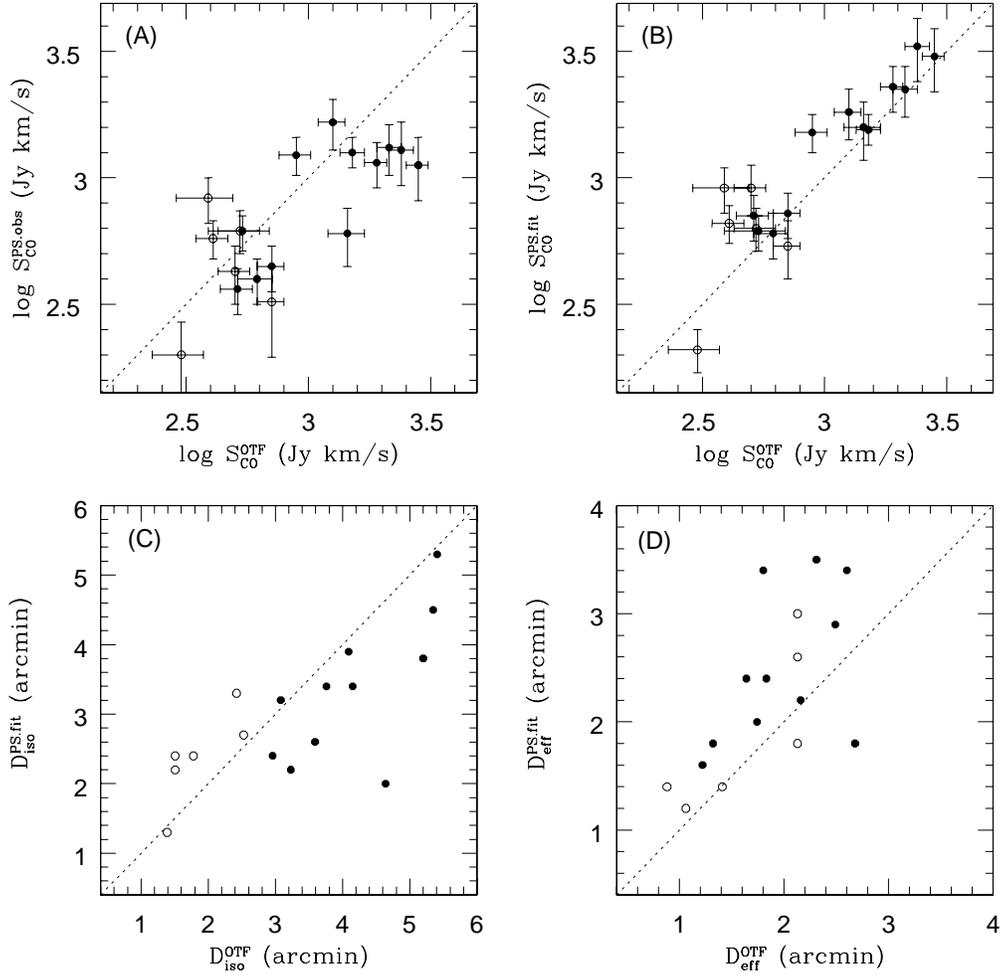} \caption{Comparison of the data
with the FCRAO extragalactic CO survey \citep{young95}. Here, ``PS''
denotes the position-switch measurements reported in the FCRAO
extragalactic CO survey data while ``OTF'' refers to our On-the-Fly
data. Panel (A) compares the total CO fluxes of the OTF measurements
with the PS measurements. Panel (B) compares the total CO fluxes
of the OTF measurements with the model fit corrected PS data. Panels
(C) and (D) compare the isophotal and effective diameters derived
from the OTF and the PS data, respectively. Isophotal diameters
${\rm D_{iso}^{OTF}}$ and ${\rm D_{iso}^{PS.fit}}$ are defined as
the diameter where the face--on CO integrated intensity falls to 1
K($\rm T_{A}^{\ast}$) km $s^{\rm -1}$. Effective diameters ${\rm
D_{eff}^{OTF}}$ and ${\rm D_{eff}^{PS.fit}}$ are the diameter which
contains 70\% of the total CO flux. The dotted line represents a
perfect linear relationship. Filled and open symbols denote Group I
and Group II galaxies, respectively.
\label{fig4}}
\end{figure*}

The FCRAO extragalactic CO survey \citep{young95} data on 300
external galaxies is the most extensive and most widely referenced
database for CO emission measurements in 300 external galaxies. A
comparison with our CO imaging survey of Virgo cluster spirals
provides an important verification of this earlier survey, which
employed sampling and modeling rather than full imaging, and %offer 
offers a quantitative constraint on the systematics resulting from the
emission modeling used by Young et al. in estimating the total CO
luminosity. Since both surveys used the same telescope and
calibration method as well as adopting the identical distance to the
Virgo Cluster, any hardware-dependent systematics are minimized.
Therefore any measured difference is narrowly constrained to the
differences in the observing methods and the modeling used by Young
et al. NGC~4536 is excluded in this comparison because the small
area coverage of our maps may have adversely affected the baseline
subtraction process.\\
% table 4 =========================================================
\clearpage
\input{tab4}
% ======================================================= end of table 4
A comparison of the FCRAO 14-m OTF and PS measurements are
summarized in Table~4.  Both the raw PS ($\rm
S_{CO}^{PS.obs}$) and the model fit $(\rm
S_{CO}^{PS.fit})$ results are listed, using the scale factor (scf)
given in the original paper \citep{young95}. An isophotal diameter
${\rm D_{iso}^{OTF}}$ is defined as the diameter where the face--on
CO integrated intensity falls to 1 K($\rm T_{A}^{\ast}$) km $s^{\rm -1}$
while an effective diameter ${\rm D_{eff}^{OTF}}$ is defined as the
diameter which contains 70\% of the total CO flux, following the
definition by Young et al.(1995).
The fractional ratio between the
OTF and PS measurements, $r$, is calculated and presented to show
the difference between the two methods of the total CO flux.
Comparisons of the OTF and PS measurements are
also shown in Figure~\ref{fig4}.\\

%================================== Figure 18
%================================== end of figure 18

The panel (A) in Figure~\ref{fig4} shows that the observed PS line
fluxes are systematically smaller than the OTF measurements, but
this is expected since ``observed PS'' data include only partial
measurements taken along the major axis.  A better agreement is
expected for galaxies with a high inclination and a small size.
And indeed galaxies with a smaller $\rm D_{25}$ and a
larger inclination appears to have a
smaller discrepancy between ${\rm S_{CO}^{PS.obs}}$ and ${\rm S_{CO}^{OTF}}$.
The model-fit measurements shown in panel (B) are more consistent with
the OTF measurements. The OTF and PS model-fit measurements of 9 
galaxies are consistent with the total CO flux to within $\rm 1 \sigma$. Out of 
the remaining 9 galaxies, 7 show larger ${\rm S_{CO}^{PS.fit}}$ than ${\rm S_{CO}^{OTF}}$ 
by a factor of 1.4 - 2.4. And, the last two galaxies (NGC 4438 and NGC 4548) 
have larger ${\rm S_{CO}^{OTF}}$ than ${\rm S_{CO}^{PS.fit}}$ by  a factor 
of 1.4. Among the 9 galaxies for which the discrepancies are larger 
than $\rm 1 \sigma$, there are 4 galaxies which disagree 
by $\geq 2 \sigma$ uncertainty, and the 4 galaxies have larger 
${\rm S_{CO}^{PS.fit}}$ than ${\rm S_{CO}^{OTF}}$ 
by  a factor of 1.6 - 2.4. We predict that peculiar CO distributions such as ring- or 
bar-like structures (e.g., Young et al. 1995) can affect the model fitting process, 
and more analysis will be done in another paper (Chung et al. 2010).\\
Panels (C) and (D) in Figure~\ref{fig4} show the comparisons of CO
diameters derived from the OTF and PS observations. For Group I
galaxies detected with high S/N ratio, isophotal diameters derived
from the OTF data (${\rm D_{iso}^{OTF}}$) are larger than those of
the PS data (${\rm D_{iso}^{PS.fit}}$). In contrast, the effective
diameters of OTF (${\rm D_{eff}^{OTF}}$) are smaller than ${\rm
D_{eff}^{PS.fit}}$.  An explanation for these apparently puzzling
trends is that CO emission is spread over a larger extent but has a
more concentrated central component in the OTF maps when compared
with the results of \citet{young95}. The effective diameters
${\rm D_{eff}^{OTF}}$ are smaller than 3 arcmin in every case, and
the isophotal diameters are located between 1/2 and 1 of the optical
diameter $D_{25}$ in most cases.
${\rm D_{iso}^{OTF}}$ for Group II galaxies are systematically
smaller than ${\rm D_{iso}^{PS.fit}}$, and this may be the result of
low S/N ratios of the OTF data.\\

\subsubsection{Comparison With the BIMA SONG Survey}

Three galaxies imaged in the OTF mode (NGC 4303, NGC 4321, \& NGC
4569) were also observed by the BIMA SONG project \citep{hel03}. The
total BIMA SONG CO fluxes reported for NGC 4303, NGC 4321, and NGC
4569 are $2427\pm145$, $2972\pm319$, and $1096\pm137$ Jy km $\rm
s^{-1}$, respectively. These integrated CO line fluxes are
systematically slightly larger ($\sim25\%$) compared with our OTF
measurements and are between the OTF and PS measurements on average.
The apparent difference may be rooted in the systematic difference
in calibration. Because of the small number of objects in common,
we can only report a broad agreement.\\

\section{DISCUSSION}

\subsection{Notes on Individual Galaxies}

In this section, we offer supplementary information on the
individual galaxies and offer comparisons with other CO, HI, and
${\rm H\alpha}$ observations. In particular, we are interested in
the environmental influence on the molecular ISM, and we make notes
of any peculiar CO distribution that may be related to the cluster
environment.\\

\subsubsection{Group I Galaxies}

\noindent \textbf{NGC 4254:}  This Sc galaxy is located 3.3$^\circ$
northwest of M87. \citet{cay90} reported that it has a very
asymmetric HI distribution with a sharp edge on the side pointing
toward M87 and an extended HI feature pointing away from M87, as if
it were compressed by the intergalactic medium. In our CO intensity
map, CO appears to extend to the south.
\citet{che06} showed an
asymmetric structure in ${\rm H\alpha}$ velocity field. Streaming
motions along the spiral arms and steep velocity rise are observed
in their atlas. Our CO velocity field map and position-velocity
diagram are in good agreement with the ${\rm H\alpha}$ kinematics.
Interferometric high resolution imaging of the central region shows
clumped CO distribution along the spiral arms, a bar-like elongation
feature, and slight depression at the
dynamical center \citep{sak99,sof03}.\\

\noindent \textbf{NGC 4302:}  This Sc galaxy is a dusty edge-on
spiral located 3.1$^\circ$ northwest of M87. Its CO distribution in
our CO intensity map consists of two molecular peaks with a central
depression. \citet{young95} modeled the CO distribution as being
offset by $0^{\prime}.45$ from its center. Our new data suggest that
this earlier study might have detected only one of the two peaks.\\

\noindent \textbf{NGC 4303:} This Sc spiral is located 8.2$^\circ$
south of M87. \citet{cay90} reported a symmetric HI distribution
with a central depression.  Our atlas shows that CO emission is
distributed along the spiral arms, and a bar-like structure
is displaced in position angle by about 20 degrees.
\citet{hel03} reached the same conclusion using the BIMA SONG survey
data. The CO emission is concentrated in the central region, unlike
the HI (see Cayatte et al. 1990), but the HI and CO velocity fields
are in good agreement. A recent burst of star formation caused by
tidal interactions with two nearby companions has been suggested by \citet{cay90}.\\

\noindent \textbf{NGC 4321:} This Sc galaxy is located 3.9$^\circ$
north of M87. Its HI emission is distributed along the spiral arms,
and the oval distortion in the inner part due to the presence of a
bar is shown from the kinematic pattern \citep{cay90}.  Our CO image
shows a well confined CO distribution on the stellar disk, plus an
asymmetric extension to the southern spiral arm. It also has an
elongated bar like structure along the major axis. The CO velocity
field shows an oval distortion in the inner part, similar to the HI
\citep{cay90}. The BIMA SONG survey detected CO at the center and
along the spiral arms \citep{hel03}, in a good agreement with our
results. \citet{sak99} reported a prominent pair of nuclear CO arms
and a sharp condensation of CO at the nucleus. \citet{che06} found
three
different pattern speeds, which is a sign of a streaming motion. \\

\noindent \textbf{NGC 4501:}  This Sbc spiral is located 2.1$^\circ$
north of M87. \citet{cay90} reported that this galaxy points its
steep HI edge toward M87, similar to NGC 4254, and suggested these
two galaxies as examples of enhanced star formation caused by
compression of the ISM.  Our CO intensity map shows a well-centered
distribution. \citet{sak99} reported a concentrated CO morphology,
and \citet{sof03} found that the north-eastern arm is much brighter
in CO than the south-western arm. The CO velocity width of NGC 4501
is one of the largest among the galaxies observed. The ${\rm
H\alpha}$ velocity field appears regular, but the
position-velocity diagram shows a complex kinematics \citep{che06}. \\

\noindent \textbf{NGC 4527:}  This Sb galaxy is located 9.8$^\circ$
south of M87.  \citet{ken88} reported that it has a uniform disk
component.  Our OTF map is consistent with the Young et al. data,
but the OTF map shows that there is a central CO concentration along
with a bar along the major axis and an asymmetric
extension to the southwest.\\

\noindent \textbf{NGC 4535:}  This SBc spiral is located 4.3$^\circ$
south of M87.   This galaxy shows an undisturbed HI distribution
with a central hole \citep{cay90}. \citet{hel03} reported that CO is
distributed at the center and along the spiral arms. \citet{sof03}
classified this galaxy as a typical single-peak type with offset
bars. Our OTF data is noisy, and the signal is very weak. The CO
intensity map shows a bar-like elongated structure in the central
region, and the position-velocity diagram shows evidence for highly
disturbed gas kinematics. \citet{che06} reported a perturbed
velocity field and streaming motions along the arms.\\

\noindent \textbf{NGC 4536:}  This Sc galaxy is located 10.2$^\circ$
south of M87.  Our map covers only a $6^{\prime} \times 4^{\prime}$
region, but the CO emission appears to be fully covered when
compared with the CO map of \citet{ken88}. CO emission is strongly
concentrated on the galactic center, and
this agrees well with the results by \citet{sof03}.\\

\noindent \textbf{NGC~4567 \& 4568:}  These two Sc galaxies are
located 1.8$^\circ$ , and they are not well separated
spatially or kinematically. \citet{cay90} suggested that the HI
emission displaced toward the south of NGC~4567 could be a sign of a
tidal interaction between the two galaxies. \citet{koo04} also
suggested ram pressure effects and a tidal interaction between the
two. However, ${\rm H\alpha}$ velocity field does not show any clear
signs of velocity disturbances \citep{che06}. Higher angular
resolution HI and CO observations by \citet{ion05} also show that
the inner gas disks show little signs of tidal disturbance, with a
symmetric bar-like or spiral-like features. Unlike many other
observed Virgo galaxies, they both show CO emission extending out to
the outer optical radii in our OTF map, including where the two disks overlap.\\

\noindent \textbf{NGC 4569:}  This Sab spiral is an HI-anemic galaxy
\citep{vdb76} located 1.7$^\circ$ northeast of M87. \citet{cay90}
reported that its HI disk is severely stripped. Interferometric CO
imaging found a symmetric CO distribution with two peaks and a
central depression \citep{hel03,sof03}. Higher resolution CO imaging
by \citet{nak05} found a highly concentrated CO distribution in the
circum-nuclear region with two off-center peaks. Our CO atlas shows
a centrally concentrated distribution with a much smaller CO radius
than the optical radius and a bar-like elongated structure across
the optical major axis. Its ${\rm H\alpha}$ velocity field is
perturbed and shows evidence for streaming motions \citep{che06}.
Our CO velocity field map shows that central velocity is shifted toward the south.\\

\noindent \textbf{NGC 4647:}  This Sc galaxy is located
3.2$^{\circ}$ east of M87 and has a companion elliptical galaxy
NGC~4649. NGC 4647 shows a slightly extended HI distribution toward
NGC 4649 \citep{cay90}. However, CO appears to point away from its
elliptical companion NGC 4649 in our intensity map. The CO
position-velocity diagram also appears disturbed.
The companion galaxy NGC~4649 is undetected in CO.\\

\noindent \textbf{NGC 4654:}  This SBc galaxy is located
3.3$^{\circ}$ northeast of M87.  The HI image shows a sharp cutoff
on the northwest side and an eastward extension, and enhanced star
formation activity is also seen in the northwest \citep{cay90}.
\citet{ken88} found an asymmetric CO distribution -- the position
45$^{\prime \prime}$ northwest of the nucleus shows a stronger line
than the central 45$^{\prime \prime}$, and this region also displays
a peak in the HI, radio continuum, and the H${\alpha}$.
\citet{sof03} found a lopsided CO distribution in the inner region
and suggested that even the nuclear region suffers from the ram
pressure effects. Our OTF CO maps also shows a clear asymmetry and
lopsidedness, but the CO peak is well centered on the stellar disk.
The lopsided CO extension is both to the northwest and to the
southwest, in the direction of M87.  The ${\rm H\alpha}$
velocity field is strongly perturbed \citep{che06}. \\

\noindent \textbf{NGC 4689:}  This Sc galaxy is located
4.3$^{\circ}$ northeast of M87. It has a large, extended HI disk
\citep{cay90}. High angular resolution CO imaging by \citet{sof03}
found a lopsided, amorphous CO morphology and without a central
peak. Our CO map shows a CO peak offset from the optical center and
extended to the south. \citet{che06} reported that ${\rm H\alpha}$ 
velocity field is slightly perturbed due to streaming associated with 
pseudo-spiral arms.\\

\subsubsection{Group II Galaxies}

\noindent \textbf{NGC 4298:}  This Sc galaxy is located 3.2$^\circ$
northwest of M87 and has a companion NGC~4302 only 2.3 arcmin away.
Stellar asymmetry due to a recent tidal interaction is seen in
optical images, and a truncated gas disk due to ram-pressure is also
suggested \citep{koo04}.  \citet{che06} suggested that its mildly
perturbed velocity field may indicate a streaming motion or a
locally warped arm. Our PVD shows stronger emission on the east
(receding) side than on the west (approaching) side, and CO emission
is probably extended toward southeast side, in the direction of its
companion galaxy NGC~4302. This feature is in good
agreement with the ${\rm H\alpha}$ kinematics discussed above.\\

\noindent \textbf{NGC 4402:}  This edge-on Sc galaxy is located
1.4$^\circ$ northwest of M87. Although the high velocity part of the
HI emission is missed by \citet{cay90}, it is obvious that the HI
disk of this galaxy is significantly smaller in extent than that of
the optical disk.  Our CO position-velocity diagram shows that CO
emission is asymmetric to one side, and \citet{young95} also
suggested the distribution model to be offset by $0^{\prime}.20$
from the center. A 10$^{\prime \prime}$ radius nuclear disk and a
more extended molecular disk with a
30$^{\prime \prime}$ diameter are found by higher angular resolution
observations \citep{sof03}.\\

\noindent \textbf{NGC 4419:}  This Sa galaxy is located 2.8$^\circ$
north of M87. \citet{ken88} proposed that NGC~4419 is on a radial
orbit which passes very close to the cluster core and strongly
interacts with the intracluster medium, resulting in a large CO/HI
flux ratio and a significant CO asymmetry. CO distribution model of
\citet{young95} shows an offset $0^{\prime}.25$ from center, and
\citet{sof03} reported that the outer molecular disk is lopsided
toward the northwest.  Our CO position-velocity diagram shows a
compact distribution
and a rapidly rising rotation speed in the central region.\\

\noindent \textbf{NGC 4438:}  This Sb galaxy is located 1.0$^\circ$
northwest of M87. \citet{cay90} reported a highly asymmetric HI
distribution with an extension pointing , and a much
smaller HI disk than the optical disk. Its ionized gas has an
off-plane filamentary morphology to the east and south of the disk
\citep{che06,ken95,ken02}. Our CO position-velocity diagram shows
that its molecular center is offset from the optical center by $\sim
0.5$ arcmin to the east. It also shows a highly disturbed structure.\\

\noindent \textbf{NGC 4548:}  This SBb galaxy is located 2.4$^\circ$
northeast of M87. This galaxy is a nearly face-on barred (SBb)
spiral which is severely HI-deficient \citep{vdb76,gio83}.
\citet{cay90} reported a ring structure in the HI distribution. The
BIMA SONG survey detected CO only at the center \citep{hel03}, and
\citet{sof03} also reported a very weak and highly concentrated CO
distribution. Our CO data is weak and too noisy to determine the molecular
distribution well. However, its global CO line profile shows moderately symmetric
double horns and CO appears to have comparatively large extent
from the integrated PVD and radial profile.\\

\noindent \textbf{NGC 4579:}  This Sab galaxy is located 1.8$^\circ$
east of M87 and is another HI-anemic spiral \citep{vdb76}. Its HI
distribution shows a ring-like structure, and it is the most
severely stripped galaxy with signs of unusual nuclear activity,
possibly fueled by gas inflow \citep{cay90}. The BIMA SONG survey
detected CO only at the center \citep{hel03}, and \citet{sof03}
reported that elongated CO distribution is displaced from the
optical bar axis by about 30 degrees. Its ${\rm H\alpha}$ velocity
field shows evidence for perturbed, streaming motions \citep{che06}.
Our CO position-velocity diagram
shows a severely disturbed morphology and kinematics. \\

\subsection{Summary of CO distribution}

Here, we summarize the CO distribution of the Virgo spiral galaxies
shown by our survey.

\begin{enumerate}
\item CO is confined to the galactic center and disk in most
galaxies. However, slight asymmetric distribution and significant
structures as well as kinematic disturbances are frequently shown.
This is consistent with previous results of CO studies (e.g., Young 
et al. 1995; Helfer et al. 2003).

\item Among the 14 galaxies which have CO intensity maps, 11 galaxies
have single CO peak center (NGC 4254, NGC 4303, NGC 4321, NGC 4501,
NGC 4527, NGC 4535, NGC 4536, NGC 4567, NGC 4568, NGC 4569, and NGC
4654). NGC 4302 has twin peaks with a central depression, and NGC
4647 and NGC 4689 have no notable CO peak at the center.

\item Five galaxies appear to have bar-like elongated structure (NGC 4303,
NGC 4321, NGC 4527, NGC 4535, and NGC 4569). Extended CO
distributions to one side are also seen in 6 galaxies (NGC 4254, NGC
4298, NGC 4527, NGC 4654, NGC 4689, and NGC 4402).

\item CO emission in NGC 4298 shows an asymmetric extension toward
the southeast, which is the direction of its companion galaxy NGC
4302. NGC 4647 shows a CO distribution pointing
away from its elliptical companion NGC 4649.
\end{enumerate}

%========================= Figure of the total Virgo CO - Fig.19
\begin{figure*}
\epsscale{1.0} \plotone{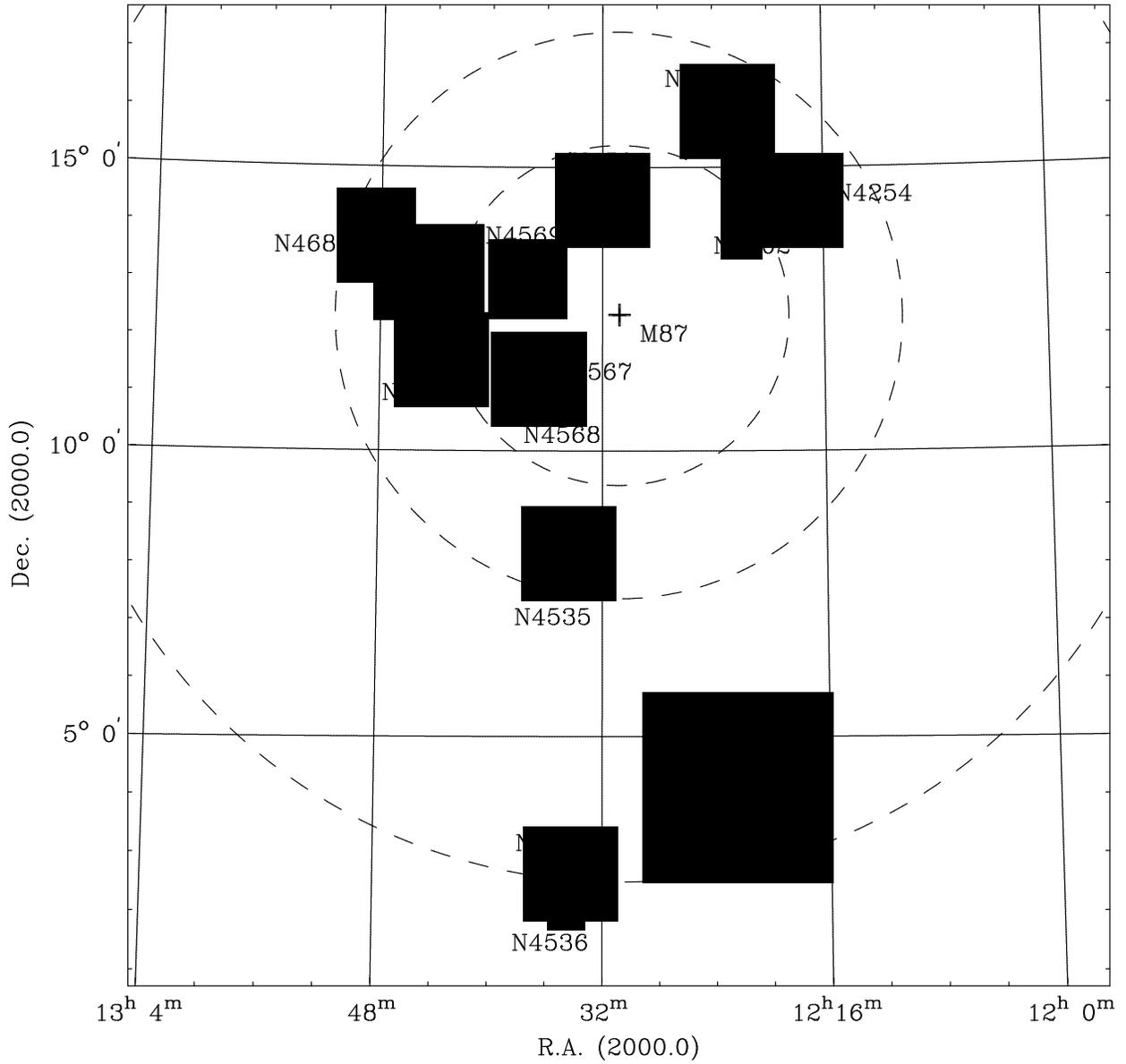} \caption{Spatial distribution of
the 14 Group I galaxies in the Virgo Cluster. Individual galaxies are
20 times enlarged to be shown more clearly. Contour levels
correspond to a H$_2$ column density of 0.75, 1.50, 2.25, and 3.00 ${\rm
10^{20}}$ ${\rm cm^{-2}}$. The radius of each
dashed circle is 3, 5, and 10 degrees, respectively.\label{fig19}}
\end{figure*}
%========================== end of figure 19

Figure~\ref{fig19} shows the total CO extent and morphology of each
of the 14 Group I galaxies at their proper position in the Virgo cluster. 
There are several noteworthy trends:

\begin{enumerate}
\item It appears that CO molecules are not strongly affected by the cluster
environments such as distance from the cluster center of M87.

\item Galaxies located in the western side of Virgo Cluster are known
to have a larger HI disk than the galaxies in the eastern
side \citep[][]{cay90}, and we find the same trend in the CO disk size
as well.

\item NGC 4298 and NGC 4647 have companion galaxies, but CO
emission is extended toward and on the opposite to its companion,
respectively.  Figure~\ref{fig19} suggests that both of their extensions
point toward the cluster center near M87.  NGC 4654 also shows a CO
extension toward M87.
\end{enumerate}

\section{SUMMARY \& CONCLUDING REMARKS}

We have carried out a $^{\rm 12}$CO(J=1--0) OTF mapping survey of 28
Virgo cluster spiral galaxies. Although the importance of molecular
contents in galaxy evolution is widely recognized, systematic CO
imaging surveys of a large galaxy sample covering the full extent of
the stellar disks %is 
are still rare.  Taking advantages of the OTF
mapping mode of the FCRAO 14-m telescope, we have imaged a large
area of each galaxy covering the entire stellar disks in a
relatively short period of time compared with the more traditional
position-switched grid-map mode.\\
We detected and mapped CO emission in 20 of the galaxies with
uniform sensitivity. Here we present their global CO properties. The
CO emission is generally well centered on the stellar disk. However,
some Virgo spirals show extended CO distribution to one side or
bar-like CO distribution. The comparison of our CO data with those of the 
FCRAO extragalactic CO survey \citep{young95} suggests that the major axis scan and
global modeling used by Young et al. was largely successful in
estimating the total CO luminosity. Young et al. has 
larger CO luminosities than our results for 7 (4) out of 18 galaxies 
by $\rm 1 \sigma$ ($\rm 2 \sigma$) uncertainty. \\
Further analysis of these data will be presented in our future
papers. Extensive multi-wavelength data are now available for
joint-analysis with our new CO data, including the HI
\citep{chung07}, radio \citep{yun01}, IR (Spitzer \& Akari), optical
(HST \& SDSS), H$\alpha$ \citep{koo04}, and UV (GALEX). We will
examine radial dependence of the star formation rate and star
formation efficiency. Their dependence on the morphological type,
luminosity, and environments will also be studied using the global
CO properties. To study environmental effects in different cluster
evolutionary stages, we will use our Virgo data with the OTF mapping
data of Ursa Major cluster spiral galaxies (Chung et al. 2010)
and single dish spectra of the Pisces filament spiral galaxies (Lee et al.
2010) obtained using the Kitt Peak 12-m telescope. We will also explore
the utility of the CO Tully-Fisher relation.\\

\acknowledgments

We thank A. Chung for valuable comments and discussions. We would also like to thank T. 
Jung and H. Kang for excellent support at the data reduction process and the system. 
Thoughtful comments from the anonymous referee are appreciated. This work was supported 
in part by NSF grants AST 0096854 and AST 0540852.

\end{document}

%% file: tab1.tex
%%
%% Begining of file `tab1.tex'

%% \documentclass{aastex}
%% \begin{document}

\begin{deluxetable}{c             % galaxy
c             % \alpha
c             % \delta
c
c             % RSA Type
c             % Type code from LEDA
c             % distance from M87
c             % Diameter at 25th mag. at B
c             % B_T
c             % T_dust
c             % inclination
c             % V_hel
} \rotate \tablecolumns{12} \tablewidth{0pc} \tablecaption{General
Properties of the Sample \label{tbl1}} \tablehead{ \colhead{Galaxy}
& \colhead{R.A.} &\colhead{Dec.} & \colhead{} &
\multicolumn{2}{c}{Type} &\colhead{$\rm R_{M87}$} &\colhead{$\rm D_{25}$}
&\colhead{$\rm B_{T}^{0}$} & \colhead{$\rm T_{dust}$}
&\colhead{i}    &\colhead{$\rm V_{hel}$} \\
\cline{2-3} \cline{5-6} \\
\colhead{}    &\colhead{(h m s.s)}    &\colhead{(d m s.s)}
 & \colhead{} &\colhead{RSA} &\colhead{LEDA} &\colhead{($^{\circ}$)}
&\colhead{(arcmin)} &\colhead{(mag)} & \colhead{(K)} &\colhead{($^{\circ}$)}
&\colhead{(\kms)} \\
\colhead{(1)} & \colhead{(2)} & \colhead{(3)} & \colhead{} &
\colhead{(4)} & \colhead{(5)} & \colhead{(6)} & \colhead{(7)}
&\colhead{(8)} & \colhead{(9)} & \colhead{(10)} & \colhead{(11)} } \startdata
N4192 & 12 13 48.28 & 14 54 01.2 & & SbII & 2.5 & 4.8 & 9.77 & 10.05 & 32.7 & 78 & -142    \\
N4212 & 12 15 39.37 & 13 54 05.7 & & Sc(s)II--III & 4.9 & 4.0 & 3.09 & 11.37 & 35.3 & 54 &  -82   \\
N4216 & 12 15 54.39 & 13 08 58.3 & & Sb(s) & 3.0 & 3.7 & 7.94 & 9.96 & 29.1 & 90 &  131    \\
N4254 & 12 18 49.56 & 14 24 59.4 & & Sc(s)I.3 & 5.2 & 3.3 & 5.37 & 10.14 & 34.9 & 29 & 2410  \\
N4293 & 12 21 12.83 & 18 22 57.6 & & Sa pec & 0.3 & 6.4 & 5.50 & 10.91 & 35.4 & 67 &   930   \\
\\
N4298 & 12 21 32.75 & 14 36 22.9 & & Sc(s)III & 5.2 & 3.2 & 2.95 & 11.46 & 29.7 & 59 & 1140 \\
N4302 & 12 21 42.41 & 14 35 52.0 & & Sc(on edge) & 5.4 & 3.1 & 4.90 & 11.05 & 29.7 & 90 & 1118   \\
N4303 & 12 21 54.89 & 04 28 25.1 & & Sc(s)I.2 & 4.0 & 8.2 & 6.17 & 10.04 & 36.3 & 19 & 1570  \\
N4321 & 12 22 54.89 & 15 49 20.7 & & Sc(s)I & 4.0 & 3.9 & 7.59 & 9.79 & 33.5 & 38 & 1579   \\
N4402 & 12 26 07.70 & 13 06 48.0 & & Sc(on edge) & 3.3 & 1.4 & 3.55 & 11.75 & 30.5 & 80 &    0  \\
\\
N4419 & 12 26 56.45 & 15 02 50.2 & & Sa & 1.1 & 2.8 & 3.31 & 11.52 & 36.1 & 82 & -254  \\
N4438 & 12 27 45.59 & 13 00 31.8 & & Sb(tides) & 0.7 & 1.0 & 8.71 & 10.52 & 32.1 & 87 &   80  \\
N4450 & 12 28 29.49 & 17 05 06.0 & & Sab pec & 2.3 & 4.7 & 5.13 & 10.61 & 28.8 & 43 & 1958  \\
N4501 & 12 31 59.16 & 14 25 13.6 & & Sbc(s)II & 3.4 & 2.1 & 6.76 & 9.67 & 31.5 & 60 & 2280   \\
N4527 & 12 34 08.50 & 02 39 10.0 & & Sb(s)II & 4.0 & 9.8 & 5.89 & 10.62 & 38.3 & 75 & 1732  \\
\\
N4535 & 12 34 20.32 & 08 11 53.8 & & SBc(s)I.3 & 5.0 & 4.3 & 6.92 & 10.35 & 34.2 & 41 & 1958   \\
N4536 & 12 34 26.93 & 02 11 18.2 & & Sc(s)I & 4.2 & 10.2 & 7.08 & 10.32 & 43.9 & 59 & 1807 \\
N4548 & 12 35 26.42 & 14 29 46.9 & & SBb(rs)I--II & 3.1 & 2.4 & 5.25 & 10.66 & 27.8 & 35 &  486  \\
N4567 & 12 36 32.71 & 11 15 28.4 & & Sc(s)II--III & 4.0 & 1.8 & 2.75 & 11.75 & 32.6 & 43 & 2265  \\
N4568 & 12 36 34.30 & 11 14 17.0 & & Sc(s)III & 4.1 & 1.8 & 4.37 & 10.95 & 32.6 & 65 & 2255   \\
\\
N4569 & 12 36 49.80 & 13 09 46.3 & & Sab(s)I--II & 2.4 & 1.7 & 10.47 & 9.59 & 34.6 & 69 & -233   \\
N4579 & 12 37 43.40 & 11 49 25.5 & & Sab(s)II & 2.8 & 1.8 & 5.62 & 10.24 & 30.7 & 39 & 1518   \\
N4647 & 12 43 32.31 & 11 34 54.7 & & Sc(rs)III & 5.2 & 3.2 & 2.82 & 11.67 & 32.0 & 34 & 1415  \\
N4649 & 12 43 39.66 & 11 33 09.4 & & SO$_{1}$(2) & -4.0 & 3.2 & 7.24 & 9.69 &   & 49 & 1139  \\
N4651 & 12 43 42.62 & 16 23 36.0 & & Sc(r)I--II & 5.2 & 5.1 & 3.98 & 11.01 & 33.4 & 50 &  804   \\
\\
N4654 & 12 43 56.57 & 13 07 35.9 & & SBc(rs)II & 5.9 & 3.3 & 5.01 & 10.56 & 33.4 & 58 & 1034  \\
N4689 & 12 47 45.60 & 13 45 46.0 & & Sc(s)II.3 & 4.7 & 4.3 & 4.57 & 11.33 & 30.5 & 39 & 1613   \\
N4710 & 12 49 38.90 & 15 09 56.7 & & SO$_{3}$(9) & -0.8 & 5.3 & 4.90 & 11.72 & 34.9 & 90 & 1325  \\
\enddata
\tablecomments{The columns are as follows. \\
Col. (1) | NGC number\\
Col. (2) \& (3) | Right Ascension and Declination at epoch J2000\\
Col. (4) | Morphological Type (RSA : Revised Shapley-Ames
Catalog)\\
Col. (5) | Numerical Morphological Type code from LEDA (Lyon/Meudon
Extragalactic Database {\rm http://leda.univ-lyon1.fr})\\
Col. (6) | Angular distance from M87 (${\rm \alpha=12h30m49.18s,\
\delta=12d23m28.9s}$)\\
Col. (7) | Major axis diameter at 25th mag arcsec{\rm$^{-2}$} in the
$B$-band (LEDA)\\
Col. (8) | Total $B$-band magnitude corrected for galactic
extinction, internal extinction, and k-correction (LEDA)\\
Col. (9) | Dust temperature from Young et al. (1989)\\
Col. (10) | Inclination (LEDA)\\
Col. (11) | Mean heliocentric radial velocity (LEDA)}
\end{deluxetable}

%% \end{document}

%%
%% End of file `tab1.tex'.

%% file: tab2.tex
%%
%% Begining of file `tab2.tex'

%% \documentclass{aastex}
%% \begin{document}

\begin{deluxetable}{
l             % galaxy
c
c             % Intensity map : I_low
c             % del_I
c
c             % V_sys
c             % del_V
c
c             % PVD : I_low
c             % del_I
c
c             % channel map : I_low
c             % del_I
} \tablecolumns{13} \tablewidth{0pc} \tablecaption{Information of CO
Atlas} \tablehead{ \colhead{Galaxy} & \colhead{} &
\multicolumn{2}{c}{CO Intensity map} & \colhead{} &
\multicolumn{2}{c}{Velocity field map}
& \colhead{} & \multicolumn{2}{c}{$\rm PVD^{a}$} & \colhead{} & \multicolumn{2}{c}{Channel map} \\
\cline{3-4} \cline{6-7} \cline{9-10} \cline{12-13} \\
\colhead{} & \colhead{} & \colhead{$\rm I_{low}$} &
\colhead{$\Delta$ I} & \colhead{} & \colhead{$\rm V_{sys}$} &
\colhead{$\Delta$ V} & \colhead{} & \colhead{$\rm I_{low}$} &
\colhead{$\Delta$ I} & \colhead{} & \colhead{$\rm I_{low}$} &
\colhead{$\Delta$ I}\\
\colhead{} & \colhead{} & \multicolumn{2}{c}{(K \kms)} & \colhead{}
& \multicolumn{2}{c}{(\kms)} & \colhead{} & \multicolumn{2}{c}{(K)}
& \colhead{} & \multicolumn{2}{c}{(K)} } \startdata
\multicolumn{3}{l}{Group I - Figure 2--14} \\
NGC 4254 & & 0.230 & 0.230  & & 2392 & 20  & & 0.020 & 0.023  & & 0.024 & 0.024 \\
NGC 4302 & & 0.096 & 0.096  & & 1147 & 40  & & 0.012 & 0.010  & & 0.016 & 0.016 \\
NGC 4303 & & 0.160 & 0.160  & & 1566 & 20  & & 0.024 & 0.024  & & 0.024 & 0.024 \\
NGC 4321 & & 0.210 & 0.210  & & 1580 & 20  & & 0.020 & 0.020  & & 0.020 & 0.020 \\
NGC 4501 & & 0.160 & 0.220  & & 2276 & 40  & & 0.020 & 0.020  & & 0.024 & 0.024 \\
NGC 4527 & & 0.240 & 0.210  & & 1736 & 40  & & 0.020 & 0.018  & & 0.021 & 0.021 \\
NGC 4535 & & 0.220 & 0.200  & & 1960 & 20  & & 0.020 & 0.020  & & 0.030 & 0.030 \\
NGC 4536 & & 0.140 & 0.140  & & 1802 & 20  & & 0.012 & 0.012  & & 0.012 & 0.012 \\
NGC 4567$^{\ast}$ & & 0.180 & 0.280  & & 2242 & 30  & & 0.018 & 0.018  & & 0.024 & 0.024 \\
NGC 4569 & & 0.200 & 0.200  & & -219 & 40  & & 0.020 & 0.020  & & 0.028 & 0.028 \\
NGC 4647 & & 0.120 & 0.120  & & 1426 & 20  & & 0.016 & 0.016  & & 0.021 & 0.021 \\
NGC 4654 & & 0.140 & 0.140  & & 1039 & 20  & & 0.020 & 0.020  & & 0.018 & 0.018 \\
NGC 4689 & & 0.120 & 0.120  & & 1621 & 20  & & 0.020 & 0.020  & & 0.023 & 0.023 \\
\\
\multicolumn{3}{l}{Group II$^{\ast\ast}$ - Figure 15--17}  \\
NGC 4298 & &  &  & &  &  & & 0.012 & 0.024 \\
NGC 4402 & &  &  & &  &  & & 0.022 & 0.033 \\
NGC 4419 & &  &  & &  &  & & 0.016 & 0.024 \\
NGC 4438 & &  &  & &  &  & & 0.018 & 0.036 \\
NGC 4548 & &  &  & &  &  & & 0.015 & 0.030 \\
NGC 4579 & &  &  & &  &  & & 0.014 & 0.021 \\
\enddata
\tablecomments{$^{\rm a}$ Position-velocity diagram (PVD) of
Group I galaxy is shown the central slice of the dataset along the
major axis, and that of Group II galaxy is the integrated PVD. 
The integration is done over some minor axis length for
each galaxy (1 or 2 beam-width for highly inclined galaxies and up to 7 beam-width
for weak CO and face-on galaxies).\\
$^{\ast}$ NGC 4567 and NGC 4568 are shown in the same map.\\
$^{\ast\ast}$ Group II galaxies have weak CO emission and do not
show much emission in the channel maps. Therefore only an integrated
PVD is shown.}
\end{deluxetable}

%% \end{document}

%%
%% End of file `tab2.tex'.

%% file: tab3.tex
%%
%% Begining of file `tab3.tex'

%% \documentclass{aastex}
%% \begin{document}

\begin{deluxetable}{
l             % galaxy
c             % rms
r@{$\pm$}l    % delW20
r@{$\pm$}l    % delW50
c             % V_sys
r@{$\pm$}l    % S_CO+-
r@{$\pm$}l    % M_H2+_
c             % D_CO^eff70
c             % D_CO^iso
} \tabletypesize{\scriptsize} \rotate \tablecolumns{13}
\tablewidth{0pc} \tablecaption{Derived CO Properties} \tablehead{
\colhead{Galaxy} & \colhead{rms} & \multicolumn{2}{c}{${W_{\rm
20}}^{a}$} & \multicolumn{2}{c}{${W_{\rm
 50}}^{b}$} & \colhead{${V_{\rm sys}}^{c}$} & \multicolumn{2}{c}{$S_{\rm
CO}$} & \multicolumn{2}{c}{$M_{\rm H_{2}}$} & \colhead{${D_{\rm
CO}^{\rm eff}}^{d}$} & \colhead{${D_{\rm CO}^{\rm iso}}^{e}$} \\
\colhead{} & \colhead{(K)} & \multicolumn{2}{c}{(\kms)} &
\multicolumn{2}{c}{(\kms)} & \colhead{(\kms)} &
\multicolumn{2}{c}{(Jy \kms)} & \multicolumn{2}{c}{(${\rm 10^{8}
M_{\odot}}$)} & \colhead{(arcmin)} & \colhead{(arcmin)} } \startdata
\multicolumn{3}{l}{Group I} & \\
NGC 4254 & 0.016 & 221 &  1 & 209 &  4 & 2392 & 2830 & 290 & 134 & 14 & 2.6 & 5.1 \\
NGC 4302 & 0.013 & 353 &  1 & 344 &  4 & 1147 &  540 & 150 &  25 &  7 & 2.7 & 3.6 \\
NGC 4303 & 0.018 & 162 &  2 & 145 &  4 & 1566 & 1920 & 200 &  90 &  9 & 2.2 & 3.9 \\
NGC 4321 & 0.021 & 239 &  2 & 217 &  4 & 1580 & 2390 & 270 & 113 & 13 & 1.8 & 4.5 \\
NGC 4501 & 0.017 & 518 &  2 & 502 &  3 & 2276 & 2130 & 240 & 101 & 11 & 2.5 & 4.8 \\
NGC 4527 & 0.017 & 376 &  2 & 358 &  5 & 1736 & 1260 & 160 &  60 &  7 & 1.7 & 3.4 \\
NGC 4535 & 0.029 & 252 &  5 & 235 &  5 & 1960 & 1450 & 250 &  68 & 12 & 2.3 & 3.9 \\
NGC 4536$^{\ast}$ & 0.010 & 318 &  2 & 267 &  5 & 1802 &  390 &  50 &  19 &  3 & 0.8 & 1.9 \\
NGC 4567$^{\ast\ast}$ & 0.018 & 297 & 3 & 272 &  4 & 2242 & 1500 & 170 &  71 &  8 &  &  \\
NGC 4569 & 0.022 & 311 &  1 & 297 &  5 & -219 &  900 & 130 & 43 &  6 & 1.3 & 2.9 \\
NGC 4647 & 0.016 & 156 &  1 & 145 & 12 & 1426 &  620 & 100 & 29 &  5 & 1.2 & 2.8 \\
NGC 4654 & 0.014 & 285 &  1 & 270 &  5 & 1039 &  710 &  90 & 33 &  4 & 1.8 & 3.4 \\
NGC 4689 & 0.020 & 182 &  1 & 171 &  3 & 1621 &  510 &  80 & 24 &  4 & 1.6 & 2.8 \\
\\
\multicolumn{3}{l}{Group II} & \\
NGC 4298 & 0.012 & 270 &  4 & 214 & 10 & 1145 &  410 &  60 & 20 &  3 & 2.1 & 3.6 \\
NGC 4402 & 0.025 & 267 &  3 & 251 &  5 &  242 &  520 & 100 & 24 &  5 & 1.4 & 1.5 \\
NGC 4419 & 0.016 & 318 &  2 & 290 & 13 & -218 &  390 & 100 & 18 &  5 & 0.9 & 2.0 \\
NGC 4438 & 0.018 & 272 &  5 & 151 &  8 &  161 &  300 &  70 & 14 &  3 & 1.1 & 2.3 \\
NGC 4548 & 0.015 & 234 &  6 & 217 &  5 &  484 &  710 &  90 & 34 &  4 & 2.1 & 3.7 \\
NGC 4579 & 0.014 & 329 &  1 & 317 &  5 & 1502 &  500 &  70 & 23 &  3 & 2.1 & 3.1 \\
\enddata
\tablecomments{$^{ a,b} $ Linewidths are corrected for the
instrumental broadening (see Eqs.~\ref{eq:ibcl2} \& \ref{eq:ibcl5}).
Uncertainties are derived by following Rhee \& van Albada (1996).\\
$^{ c} $ CO systemic velocity\\
$^{ d} $ Diameter which contains 70\% of the total CO flux\\
$^{ e} $ Diameter where the mean face--on surface density
of {\rm H$_{2}$} falls to {\rm 1 M$_{\odot}$ pc$^{-2}$}. {\rm
H$_{2}$} surface density distribution (in unit of {\rm M$_{\odot}$
pc$^{-2}$}) is derived from CO radial distribution (in unit of {\rm
K km s$^{-1}$}) assuming a constant CO to {\rm H$_{2}$} conversion
factor within a whole galaxy ($\chi = 3 \times 10^{20}$ ${\rm
cm}^{-2}$ ( K[T$_{\rm
R}$] km s$^{\rm -1}$ )$^{\rm -1}$ \citep{you91}).\\
$^{\ast} $ CO properties of NGC 4536 is derived from the datacube
mapped by $6^{\prime}\times4^{\prime}$ size.\\
$^{\ast\ast}$The entries for NGC~4567 are the total values for
NGC~4567 and NGC~4568 as a pair. CO diameters are not given
for these galaxies.}
\end{deluxetable}

%% \end{document}

%%
%% End of file `tab3.tex'.

%% file: tab4.tex
%%
%% Begining of file `tab4.tex'

%% \documentclass{aastex}
%% \begin{document}

\begin{deluxetable}{
l             % galaxy
c             %
r@{$\pm$}l    % S_CO
c             % D_iso^OTF
c             % D_eff^OTF
c
r@{$\pm$}l    % S_CO^fit of KY
c             % % obs.
c             % S_CO^obs of KY
c             % D_iso^PS.fit
c             % D_eff^PS.fit
c             %
r             % % of the differences
} \tabletypesize{\scriptsize} \rotate \tablecolumns{15}
\tablewidth{0pc} \tablecaption{Comparison with \citet{young95}}
\tablehead{ \colhead{Galaxy} & \colhead{} & \multicolumn{4}{c}{FCRAO
14-m (OTF) $^{a}$} & \colhead{} & \multicolumn{6}{c}{FCRAO 14-m
(Position Switching) $^{b}$} & \colhead{} & \colhead{} \\
\cline{3-6} \cline{8-13} \\
\colhead{} & \colhead{} & \multicolumn{2}{c}{$S_{\rm CO}^{\rm OTF}$} &
\colhead{$D_{\rm iso}^{\rm OTF}$} & \colhead{$D_{\rm eff}^{\rm OTF}$} & 
\colhead{} & \multicolumn{2}{c}{$S_{\rm CO}^{\rm PS.fit}$} & \colhead{scf} & 
\colhead{$S_{\rm CO}^{\rm PS.obs}$} & \colhead{$D_{\rm iso}^{\rm PS.fit}$} & 
\colhead{$D_{\rm eff}^{\rm PS.fit}$} & \colhead{} & \colhead{r} \\
\colhead{} & \colhead{} & \multicolumn{2}{c}{(Jy \kms)} & \colhead{(arcmin)} & \colhead{(arcmin)} & \colhead{} & 
\multicolumn{2}{c}{(Jy \kms)} & \colhead{(\%)} & \colhead{(Jy \kms)} & \colhead{(arcmin)} & \colhead{(arcmin)} & \colhead{}
& \colhead{(\%)}\\
\colhead{(1)} & \colhead{} & \multicolumn{2}{c}{(2)} & \colhead{(3)}
& \colhead{(4)} & \colhead{} & \multicolumn{2}{c}{(5)}
& \colhead{(6)} & \colhead{(7)} & \colhead{(8)} &
\colhead{(9)} & \colhead{} & \colhead{(10)} } 
\startdata
\multicolumn{3}{l}{Group I} \\
NGC 4254 & & 2830 & 290 & 5.4 & 2.6 & & 3000 & 850 &  37 & 1110 & 5.3 & 3.4 & &  6  \\
NGC 4302 & &  540 & 150 & 4.6 & 2.7 & &  620 & 100 & 100 &  620 & 2.0 & 1.8 & &  15  \\
NGC 4303 & & 1920 & 200 & 4.2 & 2.2 & & 2280 & 470 &  50 & 1140 & 3.4 & 2.2 & &  19 \\
NGC 4321 & & 2390 & 270 & 5.4 & 1.8 & & 3340 & 920 &  39 & 1300 & 4.5 & 3.4 & &  40 \\
NGC 4501 & & 2130 & 240 & 5.2 & 2.5 & & 2220 & 480 &  59 & 1310 & 3.8 & 2.9 & &   4 \\
NGC 4527 & & 1260 & 160 & 3.8 & 1.7 & & 1800 & 410 &  92 & 1660 & 3.4 & 2.0 & &  43  \\
NGC 4535 & & 1450 & 250 & 4.1 & 2.3 & & 1570 & 410 &  38 &  600 & 3.9 & 3.5 & &   8  \\
%NGC 4536 & &  390 &  50 & &  740 & 130 & 96 &  710 & &  90  & &  \\
NGC 4567 $^{\ast}$ & & 1500 & 170 &   &   & & 1550 & 210 & 81 & 1250 &   &   & & 3  \\
NGC 4569 & &  900 & 130 & 3.1 & 1.3 & & 1500 & 260 &  82 & 1230 & 3.2 & 1.8 & &  67  \\
NGC 4647 & &  620 & 100 & 3.2 & 1.2 & &  600 & 120 &  66 &  400 & 2.2 & 1.6 & &  -3  \\
NGC 4654 & &  710 &  90 & 3.6 & 1.8 & &  730 & 150 &  62 &  450 & 2.6 & 2.4 & &   3  \\
NGC 4689 & &  510 &  80 & 3.0 & 1.6 & &  710 & 150 &  50 &  360 & 2.4 & 2.4 & &  39   \\
\\
\multicolumn{3}{l}{Group II}  \\
NGC 4298 & &  410 &  60 & 1.5 & 2.1 & &  660 & 110 &  86 &  570 & 2.4 & 1.8 & &  61   \\
NGC 4402 & &  520 & 100 & 1.8 & 1.4 & &  630 & 120 &  98 &  620 & 2.4 & 1.4 & &  21  \\
NGC 4419 & &  390 & 100 & 1.5 & 0.9 & &  920 & 190 &  91 &  840 & 2.2 & 1.4 & & 136    \\
NGC 4438 & &  300 &  70 & 1.4 & 1.1 & &  210 &  40 &  94 &  200 & 1.3 & 1.2 & & -27   \\
NGC 4548 & &  710 &  90 & 2.4 & 2.1 & &  540 & 140 &  60 &  320 & 3.3 & 3.0 & & -24    \\
NGC 4579 & &  500 &  70 & 2.5 & 2.1 & &  910 & 200 &  47 &  430 & 2.7 & 2.6 & & 82    \\
\enddata
\tablecomments{$ ^{a}$ Our new OTF observation results\\
 $ ^{b}$ Position-Switching observation results by Young et
 al. (1995)\\
Col. (1) | NGC number \\
Col. (2) | Total CO flux, ${\rm S_{CO}^{OTF}}$, and its rms, ${\rm
\sigma_{OTF}}$, derived from
our OTF observations\\
Col. (3) \& (4) | The isophotal and effective CO diameter, $\rm
D_{iso}^{OTF }$ and $\rm D_{eff}^{OTF}$, derived from our OTF
observation. To compare with Young et al. (1995), $\rm
D_{iso}^{OTF}$ in here is derived at the diameter where the face--on
CO integrated intensity falls to 1 K($\rm T_{A}^{\ast}$) km $s^{\rm -1}$.\\
Col. (5) | The fitted total CO flux, ${\rm S_{CO}^{PS.fit}}$, and
its rms, ${\rm \sigma_{fit}}$, by Young et al. (1995) \\
Col. (6) | Scale factor which is the percentage of the total CO
emission sampled by the observations by Young et al. (1995) \\
Col. (7) | Recomputed Young et al. PS observed CO flux, ${\rm
S_{CO}^{PS.obs}}$, using Col. (5) and Col. (6).\\
Col. (8) \& (9) | The isophotal and effective CO diameter, $\rm
D_{iso}^{PS.fit}$ and $\rm D_{eff}^{PS.fit}$, which is the diameter
where the face--on CO integrated intensity falls to 1 K($\rm
T_{A}^{\ast}$) km $s^{\rm -1}$ and the diameter which contains 70\%
of the total CO flux for the best--fitting model,
respectively (Young et al. 1995).\\
Col. (10) | Fractional ratio of the difference between
${\rm S_{CO}^{OTF}}$ and ${\rm S_{CO}^{PS.fit}}$, calculated as
${\rm
(S_{CO}^{PS.fit}-S_{CO}^{OTF})/S_{CO}^{OTF}} \times 100$.\\
$ ^{\ast}$ NGC~4567 and NGC~4568 cannot be separated in our OTF
observations, and the summed total flux is reported. For the PS
results, the equivalent summed CO flux of NGC 4567 and NGC 4568 is
listed (Young et al. 1995). }
\end{deluxetable}

%% \end{document}

%%
%% End of file `tab4.tex'.

%% file: ms.bbl
\begin{thebibliography}{}
%\bibitem[Binggeli, Sandage, \& Tammann(1985)]{bin85} Binggeli, B., Sandage, A., \& Tammann, G. A. 1985, AJ, 90, 1981
\bibitem[Boselli \& Gavazzi(2006)]{bos06} Boselli, A., \& Gavazzi, G. 2006, PASP, 118, 517
%\bibitem[Boselli et al.(2003)]{bos03} Boselli, A., Gavazzi, G., \& Sanvito, G. 2003, A\&A, 402, 37
\bibitem[Broeils \& Rhee(1997)]{broe97} Broeils, A. H., \& Rhee, M.-H. 1997, ApJS, 111, 143
\bibitem[Brosch et al.(1997)]{bro97} Brosch, N. et al., 1997, ApJS, 111, 143
\bibitem[Cayatte et al.(1990)]{cay90} Cayatte, V. et al., 1990, AJ, 100, 604
\bibitem[Chemin et al.(2006)]{che06} Chemin, L. et al., 2006, MNRAS, 366, 812
\bibitem[Chung(2007)]{chu07} Chung, A. 2007, PhD Thesis, Columbia University
\bibitem[Chung et al.(2007)]{chung07} Chung, A., van Gorkom, J. H., Kenney, J. D. P., \& Vollmer, B. 2007, \apjl, 659, L115
\bibitem[Chung et al.(2010)]{chung10} Chung, A. et al., 2010, ApJ, in preparation
\bibitem[Chung et al.(2005)]{chu05a} Chung, E. J., Kim, H., \& Rhee, M.-H. 2005a, JKAS, 38, 17
\bibitem[Chung et al.(2005)]{chu05b} Chung, E. J., Kim, H., \& Rhee, M.-H. 2005b, JKAS, 38, 371
\bibitem[Chung et al.(2006)]{chu06} Chung, E. J., Kim, H., \& Rhee, M.-H. 2006, JASS, 23, 269
\bibitem[Chung et al.(2010)]{chu10} Chung, E. J. et al., 2010, JASS, in preparation
\bibitem[Condon et al.(1998)]{con98} Condon, J. J. et al., 1998, AJ, 115, 1693
\bibitem[Cote et al.(2004)]{cot04} Cote, P. et al., 2004, ApJS, 153, 223
\bibitem[Dale et al.(2007)]{dal07} Dale, D. A. et al., 2007, ApJ, 655, 863
\bibitem[Giovanelli \& Haynes(1983)]{gio83} Giovanelli, R., \& Haynes, M. P. 1983, AJ, 88, 881
\bibitem[Giovanelli et al.(2007)]{gio07} Giovanelli, R. et al., 2007, AJ, 133, 2569
\bibitem[Helfer et al.(2003)]{hel03} Helfer, T. T. et al., 2003, ApJS, 145, 259
\bibitem[Iono et al.(2005)]{ion05} Iono, D., Yun, M. S., \& Ho, P. T. P. 2005, ApJS, 158, 1
\bibitem[Kenney \& Young(1988)]{ken88} Kenney, J. D. P., \& Young, J. S. 1988, ApJS, 66, 261
\bibitem[Kenney \& Young(1989)]{ken89} Kenney, J. D. P., \& Young, J. S. 1989, ApJ, 344, 171
\bibitem[Kenney et al.(1995)]{ken95} Kenney, J. D. P. et al., 1995, ApJ, 438, 135
\bibitem[Kenney \& Yale(2002)]{ken02} Kenney, J. D. P., \& Yale, E. E. 2002, ApJ, 567, 865
\bibitem[Kenney et al.(2008)]{ken08} Kenney, J. D. P. et al., 2008, The Evolving ISM in the Milky Way \& Nearby Galaxies, ed. K. Sheth, A. Noriega-Crespo, J. Ingalls, \& R. Paladini, published online at heep://ssc.spitzer.caltech.edu/mtgs/ismevol/ (arXiv:0803.2532)
\bibitem[Koopmann \& Kenney(2004)]{koo04} Koopmann, R. A., \& Kenney, J. D. P. 2004, ApJ, 613, 866
\bibitem[Larson(2003)]{lar03} Larson, R. B. 2003, Rep.Prog.Phys., 66, 1651
\bibitem[Lee(2010)]{lee10} Lee, M.-Y. et al., 2010, ApJ, in preparation
\bibitem[Lucy(1974)]{luc74} Lucy, L. B. 1974, AJ, 79, 754
\bibitem[Mei et al.(2007)]{mei07} Mei, S. et al., 2007, ApJ, 655, 144
\bibitem[Nakanishi et al.(2005)]{nak05} Nakanishi, H., Sofue, Y., \& Koda, J. 2005, PASJ, 57, 905
\bibitem[Rhee(1996)]{rhee96} Rhee, M.-H. 1996, PhD Thesis, University of Groningen
\bibitem[Rhee \& Broeils(2005)]{rhe05} Rhee, M.-H., \& Broeils, A. H. 2005, JASS, 22, 89
\bibitem[Rhee \& van Albada(1996)]{rhe96} Rhee, M.-H., \& van Albada, T. S. 1996, A\&AS, 115, 407
\bibitem[Sakamoto et al.(1999)]{sak99} Sakamoto, K. et al., 1999, ApJS, 124, 403
\bibitem[Sandage \& Tammann(2006)]{san06} Sandage, A., \& Tammann, G. A. 2006, ApJ, submitted (astro-ph/0608677)
\bibitem[Sofue et al.(2003)]{sof03} Sofue, Y. et al., 2003, PASJ, 55, 17S
\bibitem[Solomon \& Barrett(1991)]{sol91} Solomon, P. M., \& Barrett, J. W. 1991, in IAU Symp. 146, Dynamics of 
Galaxies and Their Molecular Cloud Distributions, ed. F. Combes \& F. Casoli (Dordrecht: Kluwer), 235
\bibitem[van den Bergh(1976)]{vdb76} van den Bergh, S. 1976, ApJ, 206, 883
\bibitem[van der Hulst et al.(1992)]{vdh92} van der Hulst, J. M. et al., 1992, in ASP Conf. Ser. 25 Astronomical Data Analysis Software and System I, ed. D. M. Worall, C. Biemesderfer, \& J. Barnes, 131
\bibitem[van Gorkom et al.(1984)]{vgo84} van Gorkom, J. H. et al., 1984, in Groups and Clusters of Galaxies, ed. F.  Mardirossian, G. Giuricin, \& M. Mezzetti (Dordrecht: Reidel), 261
\bibitem[Verheijen(1997)]{ver97} Verheijen, M. A. W. 1997, PhD Thesis, University of Groningen
\bibitem[Warmels(1988)]{war88} Warmels, R. H. 1988, A\&AS, 72, 427
\bibitem[Yasuda et al.(1997)]{yas97} Yasuda, N. et al., 1997, ApJS, 108, 417
\bibitem[Young et al.(1989)]{young89} Young, J. S. et al., 1989, ApJS, 70, 699 %in table 1
\bibitem[Young \& Scoville(1991)]{you91} Young, J. S., \& Scoville, N. 1991, ARA\&A, 29, 581
\bibitem[Young et al.(1995)]{young95} Young, J. S. et al., 1995, ApJS, 98, 219
\bibitem[Yun, Reddy, \& Condon(2001)]{yun01} Yun, M. S., Reddy, N. A., \& Condon, J. J. 2001, ApJ, 554, 803
\end{thebibliography}
